\documentclass[final]{agujournal2019}
\usepackage{url} %this package should fix any errors with URLs in refs.
\usepackage{lineno}
\usepackage{amsmath,amsfonts,amssymb}
\usepackage[inline]{trackchanges} %for better track changes. finalnew option will compile document with changes incorporated.
\usepackage{soul}
\usepackage[normalem]{ulem}
\newcommand*\patchAmsMathEnvironmentForLineno[1]{
  \expandafter\let\csname old#1\expandafter\endcsname\csname #1\endcsname
  \expandafter\let\csname oldend#1\expandafter\endcsname\csname end#1\endcsname
  \renewenvironment{#1}
  {\linenomath\csname old#1\endcsname}
  {\csname oldend#1\endcsname\endlinenomath}}
  \newcommand*\patchBothAmsMathEnvironmentsForLineno[1]{
  \patchAmsMathEnvironmentForLineno{#1}
  \patchAmsMathEnvironmentForLineno{#1*}}
  \AtBeginDocument{
  \patchBothAmsMathEnvironmentsForLineno{equation}
  \patchBothAmsMathEnvironmentsForLineno{align}
  \patchBothAmsMathEnvironmentsForLineno{flalign}
  \patchBothAmsMathEnvironmentsForLineno{alignat}
  \patchBothAmsMathEnvironmentsForLineno{gather}
  \patchBothAmsMathEnvironmentsForLineno{multline}
}

\newcommand{\mycomment}[1]{}
\draftfalse

\journalname{Journal of Advances in Modeling Earth Systems (JAMES)}

\begin{document}

%% ------------------------------------------------------------------------ %%
%  Title
%
% (A title should be specific, informative, and brief. Use
% abbreviations only if they are defined in the abstract. Titles that
% start with general keywords then specific terms are optimized in
% searches)
%
%% ------------------------------------------------------------------------ %%

% Example: \title{This is a test title}

%\title{Data-driven stochastic parameterizations of subgrid mesoscale eddies in an idealized ocean model}
\title{Generative data-driven approaches for stochastic subgrid parameterizations in an idealized ocean model}

%% ------------------------------------------------------------------------ %%
%
%  AUTHORS AND AFFILIATIONS
%
%% ------------------------------------------------------------------------ %%

% Authors are individuals who have significantly contributed to the
% research and preparation of the article. Group authors are allowed, if
% each author in the group is separately identified in an appendix.)

% List authors by first name or initial followed by last name and
% separated by commas. Use \affil{} to number affiliations, and
% \thanks{} for author notes.
% Additional author notes should be indicated with \thanks{} (for
% example, for current addresses).

% Example: \authors{A. B. Author\affil{1}\thanks{Current address, Antartica}, B. C. Author\affil{2,3}, and D. E.
% Author\affil{3,4}\thanks{Also funded by Monsanto.}}

\authors{Pavel Perezhogin\affil{1}, Laure Zanna\affil{1}, Carlos Fernandez-Granda\affil{1,2}}

% \affiliation{1}{First Affiliation}
% \affiliation{2}{Second Affiliation}
% \affiliation{3}{Third Affiliation}
% \affiliation{4}{Fourth Affiliation}

\affiliation{1}{Courant Institute of Mathematical Sciences, New York University, New York, NY, USA}
\affiliation{2}{Center for Data Science, New York University, New York, NY, USA}

%% Corresponding Author:
% Corresponding author mailing address and e-mail address:

% (include name and email addresses of the corresponding author.  More
% than one corresponding author is allowed in this LaTeX file and for
% publication; but only one corresponding author is allowed in our
% editorial system.)

% Example: \correspondingauthor{First and Last Name}{email@address.edu}

\correspondingauthor{P.A. Perezhogin}{pperezhogin@gmail.com}

%% Keypoints, final entry on title page.

%  List up to three key points (at least one is required)
%  Key Points summarize the main points and conclusions of the article
%  Each must be 140 characters or fewer with no special characters or punctuation and must be complete sentences

% Example:
% \begin{keypoints}
% \item	List up to three key points (at least one is required)
% \item	Key Points summarize the main points and conclusions of the article
% \item	Each must be 140 characters or fewer with no special characters or punctuation and must be complete sentences
% \end{keypoints}

\begin{keypoints}
\item We propose generative machine learning (ML) models to build stochastic parameterization of subgrid mesoscale eddies
\item Generative models produce a flow-dependent estimation of the uncertainty with spatially correlated stochastic forcing
\item Generative models demonstrate superior numerical stability and outperform baseline ML models in online simulations at the coarsest grid
%\item The proposed parameterizations are the most efficient at the coarsest resolution and improve tails of PDFs of turbulence fields
%\item Generative models demonstrate improved numerical stability properties compared to baseline deterministic and stochastic models
%\item Probable keypoint or conclusion: we suggest new loss functions to train subgrid models?
\end{keypoints}

%% ------------------------------------------------------------------------ %%
%
%  ABSTRACT and PLAIN LANGUAGE SUMMARY
%
% A good Abstract will begin with a short description of the problem
% being addressed, briefly describe the new data or analyses, then
% briefly states the main conclusion(s) and how they are supported and
% uncertainties.

% The Plain Language Summary should be written for a broad audience,
% including journalists and the science-interested public, that will not have 
% a background in your field.
%
% A Plain Language Summary is required in GRL, JGR: Planets, JGR: Biogeosciences,
% JGR: Oceans, G-Cubed, Reviews of Geophysics, and JAMES.
% see http://sharingscience.agu.org/creating-plain-language-summary/)
%
%% ------------------------------------------------------------------------ %%

%% \begin{abstract} starts the second page

\begin{abstract}

Subgrid parameterizations of mesoscale eddies continue to be in demand for climate simulations. These subgrid parameterizations can be powerfully designed using physics and/or data-driven methods, with uncertainty quantification. For example, \citeA{guillaumin2021stochastic} proposed a Machine Learning (ML) model that predicts subgrid forcing and its local uncertainty. The major assumption and potential drawback of this model is the statistical independence of stochastic residuals between grid points. Here, we aim to improve the simulation of stochastic forcing with generative models of ML, such as Generative adversarial network (GAN) and Variational autoencoder (VAE). Generative models learn the distribution of subgrid forcing conditioned on the resolved flow directly from data and they can produce new samples from this distribution. Generative models can potentially capture not only the spatial correlation but any statistically significant property of subgrid forcing. We test the proposed stochastic parameterizations offline and online in an idealized ocean model.  We show that generative models are able to predict subgrid forcing and its uncertainty with spatially correlated stochastic forcing. Online simulations for a range of resolutions demonstrated that generative models are superior to the baseline ML model at the coarsest resolution.
\end{abstract}

\section*{Plain Language Summary}
The climate system includes physical phenomena on a wide range of scales from millimeter scale in the boundary layer to planetary scale.  Numerical models used for climate projections can directly simulate only the largest spatiotemporal scales of the flow, while the missing physics due to unresolved (or subgrid) flows must be parameterized. The prediction of the missing term given only the information about the resolved flow is a difficult task, given in part the uncertainty associated with the state of the unresolved eddies which were discarded. Generative machine learning models have demonstrated exceptional ability to create realistic images obeying complex distributions learned directly from data. In this work, we leverage the generative machine learning approach to build a stochastic parameterization of the subgrid eddies which is able to sample many possible realizations of the missing physics forcing. The new stochastic models have shown excellent performance in predicting the missing physics term and have the promise to improve the simulation of turbulence when implemented online in the idealized ocean model.

\section{Introduction}
Mesoscale eddies, with a horizontal scale roughly equal to the Rossby deformation radius, play a crucial role in ocean circulation. Mesoscale eddies carry most of the kinetic energy in the ocean and account for a substantial part of the transport of momentum, heat, and salt \cite{vallis2017atmospheric}. The dynamics of mesoscale eddies involve a variety of complex physical processes: potential to kinetic energy conversion, upscale energy transfer, upgradient fluxes, sharpening of jet currents, along-isopycnal mixing and bolus advection. Primitive equations can potentially capture all these processes if all the relevant spatial scales of motion are directly resolved on the computational grid. However, direct simulation of mesoscale eddies remains computationally expensive, especially in high latitudes where the deformation radius decreases \cite{hewitt2020resolving}. 

Modern global ocean models have an eddy-permitting resolution (around $1/4^o$, \citeA{haarsma2016high}), such that the largest mesoscale eddies are resolved but smaller ones are not; therefore the effect of these smaller unresolved (subgrid) mesoscale eddies is missing and needs to be parameterized.
%that is largest mesoscale eddies are represented on the grid, but an effect of unresolved (subgrid) mesoscale eddies is a missing physical process that needs to be parameterized.
A range of grid resolutions where a physical process is partially (but not fully) resolved is often referred to as the gray zone \cite{berner2017stochastic, christensen2022parametrization}. 
Traditional methods to parameterize mesoscale eddies \cite{redi1982oceanic, gent1990isopycnal} were designed to describe their mean effect on the large-scale flow. These parameterizations are suitable for ocean models with a very coarse horizontal resolution, where there is an approximate scale separation between the grid step and the size of mesoscale eddies, but not  for the gray zone. 
%In a gray zone, the dynamics of mesoscale eddies need to be split into resolved and subgrid contributions depending on the grid step. 

The "Large eddy simulation" approach (LES, \citeA{fox2008can, sagaut2006large}) is a technique to build a mesoscale eddy parameterization in the gray zone. The LES framework introduces a spatial filtering (and coarse-graining) operator which splits the flow into resolved and subgrid components. 
The filter mimics the effect of finite resolution and its width is proportional to the grid step of the coarse model. The effect of subgrid eddies on the resolved flow is referred to as a subgrid forcing and is diagnosed from the output of the high-resolution model by applying the spatial filter to the governing equations. A subgrid model or parameterization is a model which relates the subgrid forcing to the resolved flow. %A subgrid model is then evaluated offline and online, also known in LES literature as a priori and a posteriori, respectively. 
In recent years many new mesoscale eddy parameterizations were proposed to better capture the effects of mesoscale eddies in the gray zone using some heuristic (or empirical) physical arguments \cite{thuburn2014cascades,jansen2014parameterizing, mana2014toward, zanna2017scale, bachman2017scale, pearson2017evaluation, bachman2018relationship, jansen2019toward, bachman2019gm, grooms2015numerical, berloff2018dynamically, juricke2020ocean}.

Machine Learning (ML) methods have recently gained traction as a new direction for developing subgrid eddy parameterizations in geophysics and turbulence \cite{rasp2018deep, Bolton2019,maulik2019subgrid, beck2019deep, yuval2020stable, guan2022stable, beucler2021climate, shamekh2022implicit, wang2022non}. ML parameterizations capture the effect of subgrid eddies on the resolved flow by \emph{training} a model in a data-driven fashion. 
The most popular approach to train ML subgrid models is to minimize the mean squared error (MSE) between their output and a subgrid forcing obtained by reducing the resolution of a high-resolution model via filtering and coarse-graining \cite{Bolton2019}. Such models typically have excellent \emph{offline} performance: they are able to accurately predict the subgrid forcing. However, the ultimate goal of subgrid parameterizations is to improve \emph{online} performance, once the parameterization is included into the coarse ocean model and the model is integrated for a long time. The coarse parameterized model should then reproduce the statistical properties of the coarse-grained high-resolution model \cite{sagaut2006large}.
% Consider the most popular approach to train an ML subgrid model -- the minimization of the mean squared error (MSE):
% \begin{equation}
%     \sum (S-\widetilde{S})^2, \label{eq:MSE_intro}
% \end{equation}
% where $S$ and $\widetilde{S}$ are subgrid forcing and subgrid model, respectively. MSE loss turned out to be a poor proxy of the online performance: 
Recent work has shown that the offline and online performance of subgrid parameterizations correlate poorly \cite{ross2022benchmarking}: models trained with the offline MSE loss may be unstable when applied online \cite{beck2019deep, maulik2019subgrid} and physically-based parameterizations have very low offline MSE but perform reasonably well online \cite{ross2022benchmarking}.
%Recently there were attempts to improve this naive approach of minimization of the MSE loss function. 
Several approaches have been proposed to improve ML parameterizations. 
\citeA{kochkov2021machine} and \citeA{frezat2022posteriori} proposed an online training procedure that improves numerical stability properties but requires a differential model and has a considerable computational cost.  \citeA{guan2022stable} suggested gradually enlarging the training dataset until the rare events in subgrid forcing are well captured. In \citeA{guan2022learning} the MSE loss function was modified with an additional constraint involving energy exchange. \citeA{frezat2021physical, guan2022learning,pawar2022frame} proposed to account for physical invariances of subgrid forcing.

%Another challenge, beyond stability and metrics, is that a possible source of model error can arise since 
Conventional subgrid parameterizations are deterministic and predict a single subgrid forcing for a given input  \cite{berner2017stochastic}, which represents the mean or most likely prediction given the resolved flow. 
However, many possible states of the subgrid eddies are typically consistent with a given resolved flow, %. In the gray zone, the grid box is no longer large enough and only a few subgrid eddies fit into the grid box. It results in an increased variability and uncertainty 
so there is inherent uncertainty in the subgrid fluxes \cite{gerard2007integrated, berner2017stochastic, christensen2022parametrization}. 
Quantifying this uncertainty requires characterizing the distribution of the subgrid forcing, conditioned on the resolved variables. The stochastic ML model of \citeA{guillaumin2021stochastic} performs uncertainty quantification by estimating the pointwise conditional mean and conditional variance of the subgrid forcing, but does not take into account spatial correlations. %A substantial drawback of GZ model compared to a full uncertainty given by $\rho(S|\overline{q})$ is the lack of spatial correlation of stochastic residuals.

%This motivates performing \emph{uncertainty quantification}, to characterize the uncertainty associated to the estimate of the subgrid forcing. 

% By analyzing high-resolution data, \citeA{palmer2012towards, arnold2013stochastic, mana2014toward, christensen2020constraining} showed that uncertainty in subgrid forcing can be represented as a conditional probability density function (cPDF), where PDF is unique for every state of the resolved variables in a grid box. \citeA{palmer2001nonlinear} suggested to account for model uncertainty in non-local way, that is cPDF should be conditioned on the resolved variables in all grid boxes.
% %That is PDF of subgrid forcing in a given grid box should be constrained on the resolved flow in all grid boxes. 
% Mathematically non-local cPDF is defined as
% \begin{equation}
%     \rho(S|\overline{q}), \label{eq:cPDF_intro}
% \end{equation}
% where $S$ is the subgrid forcing, $\overline{q}$ is a resolved flow and we consider $S$ and $\overline{q}$ as 2D or 3D fields. We refer to a model predicting $\rho(S|\overline{q})$ as an uncertainty quantification (UQ) model. Prediction of cPDF may be seen as an alternative offline training method in comparison to the minimization of the MSE loss.
% We use UQ model to build a stochastic parameterization that is implemented into the coarse ocean model and evaluated online. Implementation does not require to know the full cPDF and it is enough to produce one sample from cPDF at every time step.

Subgrid models incorporating uncertainty quantification (UQ) can be used to build \emph{stochastic} parameterizations, where the subgrid forcing is random. Stochastic parameterizations are widely used in climate models and have been shown to improve the mean state and variability \cite{palmer2000predicting, berner2012systematic, berner2017stochastic, christensen2017stochastic, juricke2017stochastic}. The two simplest stochastic parameterizations are Stochastically perturbed parameterization tendency (SPPT, \citeA{buizza1999stochastic, andrejczuk2016oceanic, subramanian2019stochastic}) which multiplies a deterministic subgrid model by a random number with unit mean and non-zero spread and Stochastic kinetic energy backscatter scheme (SKEBS, \citeA{berner2009spectral, storto2021new}) which introduces additive stochastic forcing. 
% Recently, more advanced ML methods were proposed to account for the uncertainty of subgrid forcing \cite{guillaumin2021stochastic, gagne2020machine, alcala2021subgrid, nadiga2022stochastic, bhouri2022history}. 
The effect of stochastic parameterizations on  online performance depends in a complex way on the associated UQ model. There is sensitivity to spatial \cite{grooms2015numerical} and temporal \cite{wilks2005effects, arnold2013stochastic, schumann1995stochastic, berner2009spectral} correlations of stochastic forcing, its non-Gaussian distribution \cite{mana2014toward, zanna2017scale} and its dependence on the resolved flow (multiplicative noise, \citeA{sura2005multiplicative, arnold2013stochastic, zacharuk2018stochastic}). %However, it becomes increasingly difficult to propose the structure of UQ model and corresponding sampling algorithm in a general case of non-stationary \cite{chasnov1991simulation} and non-homogeneous \cite{berloff2005random} turbulence.
%Usually, data-driven models predict only a few properties of the conditional distribution $\rho(S|\overline{q})$. The optimal deterministic model minimizing the MSE loss approximates the mean of $\rho(S|\overline{q})$ \cite{bishop2006pattern}. The stochastic model of \citeA{guillaumin2021stochastic} referred to as GZ predicts spatial fields of the first two moments of $\rho(S|\overline{q})$: conditional mean and conditional variance. A substantial drawback of GZ model compared to a full uncertainty given by $\rho(S|\overline{q})$ is the lack of spatial correlation of stochastic residuals.

In this work, we propose to leverage two powerful uncertainty-quantification ML frameworks to data-driven subgrid parameterization of mesoscale eddies: variational autoencoder (VAEs, \citeA{kingma2013auto}) and generative 
adversarial networks (GANs, \citeA{goodfellow2014generative}). These frameworks provide a data-driven characterization of the conditional distribution of the subgrid forcing given the resolved flow. The resulting ML models are \emph{generative}, meaning that they allow us to sample from the conditional distribution, and can be therefore directly deployed as stochastic parameterizations. 
%can potentially improve UQ of the subgrid forcing \cite{gagne2020machine, alcala2021subgrid, nadiga2022stochastic}. The aim of generative models is to learn the conditional probability distribution $\rho(S|\overline{q})$ directly from data. Instead of learning the analytical form of the density $\rho(S|\overline{q})$, they learn how to sample from this distribution, and thus it is convenient to use a generative model as a stochastic parameterization.
Our proposed ML models do not contain a-priori assumptions about the structure of the statistical model.  These ML models can therefore potentially capture any statistically significant properties of the subgrid forcing such as the spatial correlation of stochastic residuals, dependence on the resolved flow, or probability distribution \cite{adler2018deep,gagne2020machine, alcala2021subgrid, nadiga2022stochastic}. In addition, generative models can be trained and tested using the same datasets, as MSE-based ML models. 

%\cite{adler2018deep, gagne2020machine, alcala2021subgrid, nadiga2022stochastic} in a range of applications. 

%Our proposed ML models do not contain assumptions about the structure of the statistical model and thus can potentially capture statistical properties of the subgrid forcing such as the spatial correlation of stochastic residuals, dependence on the resolved flow, or probability distribution \cite{adler2018deep}. In addition, they can be trained offline using the same data as in MSE-based offline ML models \cite{adler2018deep}. \cite{gagne2020machine, alcala2021subgrid, nadiga2022stochastic}

%In this work, we use generative models GAN (Generative adversarial network, \citeA{goodfellow2014generative}) and VAE (Variational autoencoder, \citeA{kingma2013auto}) to build a subgrid parameterization of mesoscale eddies accounting for the uncertainty in subgrid forcing. 
We implement our generative models in an idealized ocean simulation and evaluate them both offline and online. 
%We train generative models on the same dataset and use similar architecture of the neural network for the inference as GZ model. Consequently, the main difference between parameterizations consists in the loss function used for training. 
Our offline analysis shows that the generative models provide a flow-dependent prediction of uncertainty. The resulting stochastic residuals are correlated in space and reproduce stochastic backscatter \cite{leslie1979application, chasnov1991simulation, frederiksen1997eddy} in the correct band of scales. Additionally, generative models accurately simulate large-scale kinetic energy backscatter \cite{thuburn2014cascades, jansen2014parameterizing} and properly energize the flow. Our online analysis shows that the generative models have better numerical stability and metrics than the baseline ML model in \citeA{guillaumin2021stochastic} at coarse resolutions.

\section{Idealized ocean model and subgrid eddy forcing} \label{sec:ocean_model}

\begin{figure}
\includegraphics[width=\textwidth]{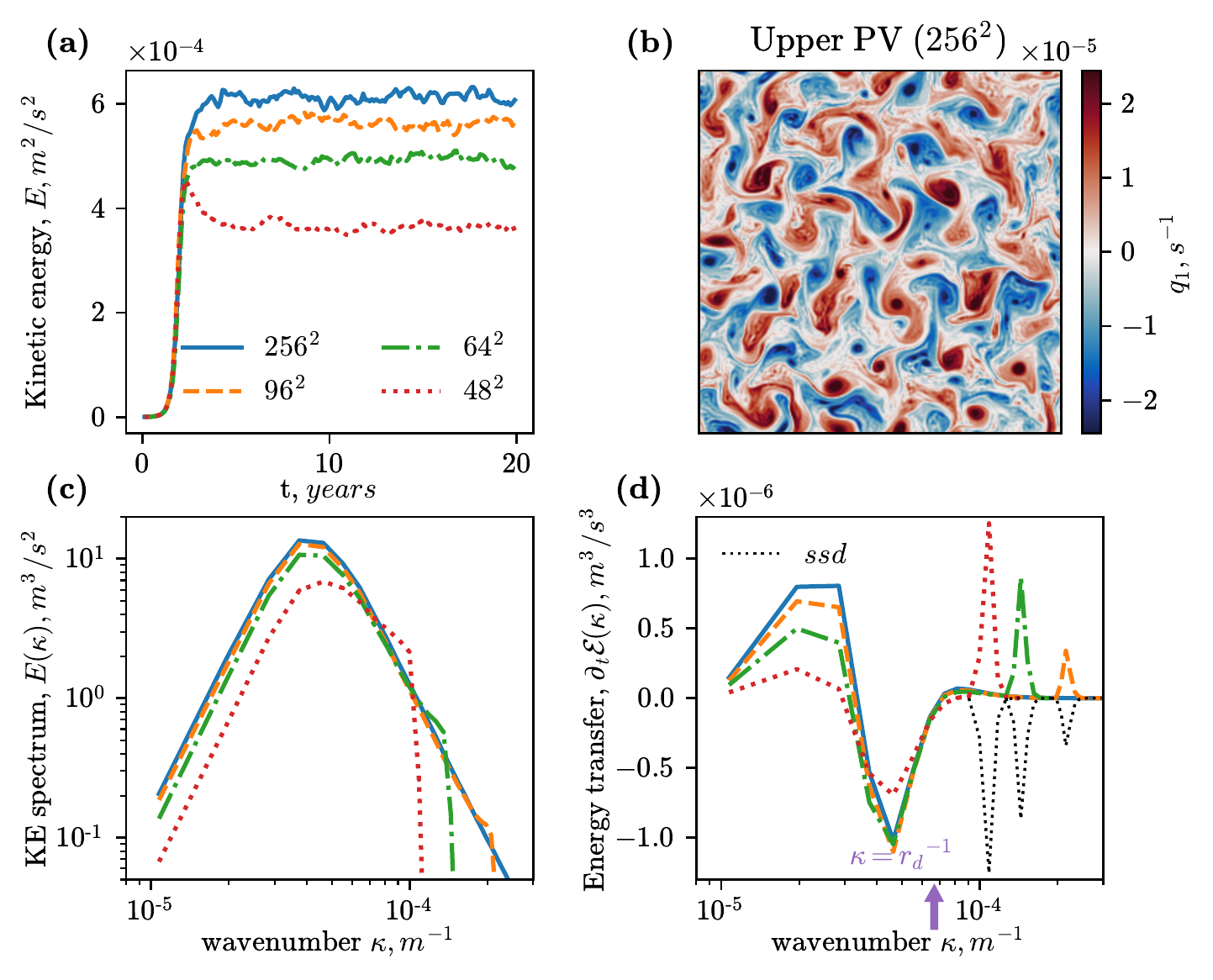}
\caption{Reference simulations at four different resolutions: (a) kinetic energy (Eq.~\eqref{eq:KE_definition}) as a function of time, (b) snapshot of the potential vorticity in the model with the finest resolution, (c) the spectral density of kinetic energy normalized as $E=\int E(\kappa) d\kappa$,
%as a function of the radial wavenumber $\kappa$, 
(d) total energy transfer from nonlinear advection (see Eq. \eqref{eq:energy_balance_unfiltered}) normalized as $\frac{\partial}{\partial t} \mathcal{E}=\int \partial_t \mathcal{E}(\kappa)d\kappa$.
%, where the total energy is defined by Eq. \eqref{eq:E_definition}. 
Coarse models fail to reproduce the energy cycle when their resolution is insufficient to resolve the deformation radius $\kappa= r_d^{-1}$ (see arrow).
Black dotted lines in panel (d) show the dissipation model ($ssd$), which causes a spurious forward energy cascade (spikes in energy transfer).}
\label{fig:fig1}
\end{figure}

In this section, we describe an idealized numerical ocean model based on quasi-geostrophic (QG) equations of layered fluid written in Python (pyqg, \citeA{pyqg}), see Figure \ref{fig:fig1}. The configuration of the QG model and the corresponding definition of subgrid forcing are similar to those in \citeA{ross2022benchmarking}. We use this model to perform offline and online evaluation of the proposed methodology to build subgrid parameterization for a range of resolutions. 
%The output of this model is used to compute nonlinear interaction between resolved and subgrid eddies, i.e. subgrid forcing. The same model, at lower horizontal resolution, is used for online simulations with the proposed subgrid parameterizations. 
% However, in this study,  we investigate a range of resolutions of the coarse model.

\subsection{Governing equations}
We solve numerically the QG equations for potential vorticity (PV) anomalies relative to the mean flow given by a prescribed vertical shear that plays the role of external forcing driving turbulence.

The two-layer QG equations in Cartesian coordinates ($x$ is zonal, $y$ is meridional) are:
\begin{gather}
    \partial_t q_m + \nabla \cdot (\mathbf{u}_m q_m) + \beta_m \partial_x \psi_m + U_m \partial_x q_m = - \delta_{m,2} r_{ek} \nabla^2 \psi_m + ssd \circ q_m, \label{eq:gov_eq_1} \\
    q_m = \nabla^2 \psi_m + (-1)^m \frac{f_0^2}{g' H_m} (\psi_1 - \psi_2), ~ m \in \{1,2\}
    \label{eq:gov_eq_2}
\end{gather}
where $m$ is the index of the fluid layer ($1$ for the upper layer and $2$ for the lower layer); $q_m$ is the potential vorticity (PV) which is conserved on Lagrangian trajectories in absence of forcing and dissipation; $\psi_m$ is the streamfunction, related to velocity as $\mathbf{u}_m=(u_m, v_m)=(-\partial_y \psi_m, \partial_x \psi_m)$; $U_m$ is the prescribed mean zonal flow (in the $x$ direction); $\beta_m = \beta + (-1)^{m+1} \frac{f_0^2}{g'H_m} (U_1-U_2)$ is the meridional gradient of potential vorticity due to differential rotation (in $\beta$-plane approximation) and prescribed mean flow; $r_{ek}$ is the bottom drag coefficient; $\delta_{m,2}$ is a Kroneker delta which indicates that drag is applied only to the lower layer; $f_0$ is the reference Coriolis frequency; $g'$ is the reduced gravity and $H_m$ is the fluid layer thickness,  $H=H_1+H_2$ is the total depth; $\nabla=(\partial_x, \partial_y)$ is a horizontal Nabla operator, where $\partial_x, \partial_y$ are partial derivatives w.r.t. $x,y$. The numerical schemes and how the small-scale dissipation ($ssd$) is applied to the governing equations are described in \ref{appendix:numerical_schemes}. The kinetic and total energy per unit mass are respectively given by:
\begin{eqnarray}
E=\frac{1}{2H} \sum_{m=1}^{2} H_m \langle |\mathbf{u}_m|^2 \rangle \label{eq:KE_definition} \\
\mathcal{E} = -\frac{1}{2H} \sum_{m=1}^{2} H_m  \langle \psi_m q_m \rangle \label{eq:E_definition} 
\end{eqnarray}
where $\langle \cdot \rangle$ is 2D spatial averaging. The QG system, described by Eq. \eqref{eq:gov_eq_1} and
\eqref{eq:gov_eq_2} is initially perturbed from rest with random noise in the upper PV field, with a subsequent evolution over the next 2--5 years exhibiting a transition to turbulence. %which is followed by the transition to turbulence within the next 2-5 years.
The initial random perturbations are limited to the range of scales of the coarsest model, and it allows to simulate similar energy growth in the transition from laminar to turbulent regimes at different grid resolutions, see Figure \ref{fig:fig1}(a). Model parameters are given in Table \ref{tab:qg_parameters} and correspond to the "eddy" configuration in \citeA{ross2022benchmarking}.

\begin{table}
\caption{Parameters of the quasi-geostrophic model in online simulations (eddy configuration from \citeA{ross2022benchmarking}).}
\centering
\begin{tabular}{lccc}
Common parameters \\
\hline
Integration time & \multicolumn{3}{c}{20 years} \\
Ensemble size & \multicolumn{3}{c}{10 runs} \\
Domain size ($L\times W$) & \multicolumn{3}{c}{1000km $\times$ 1000km} \\
Boundary conditions & \multicolumn{3}{c}{periodic} \\
Ocean depth ($H=H_1+H_2$) & \multicolumn{3}{c}{2500m} \\
Upper layer thickness ($H_1$) & \multicolumn{3}{c}{500m} \\
Bottom drag ($r_{ek}$) & \multicolumn{3}{c}{$5.787 \cdot 10^{-7} \mathrm{s}^{-1}$} \\
Differential rotation ($\beta$) & \multicolumn{3}{c}{$1.5 \cdot 10^{-11} (\mathrm{m~s})^{-1}$} \\
Deformation radius ($r_d = \frac{g'}{f_0^2} \frac{H_1 H_2}{H}$) & \multicolumn{3}{c}{15km} \\
Mean flow ($U_1, U_2$) & \multicolumn{3}{c}{(0.025m/s, 0m/s)} \\
Velocity scale ($\sqrt{2E}$) & \multicolumn{3}{c}{$\approx$ 0.035m/s} \\
\\
Grid parameters & resolution & grid step ($\Delta x$) & time step ($\Delta t$) \\
\hline
High resolution & $256\times 256$ & 3.9km & 1hour \\
\hline
Coarse models & $96\times 96$ & 10.4km & 2hour \\
& $64\times 64$ & 15.6km & 4hour \\
& $48\times 48$ & 20.8km & 1,2,4,8 hour \\
\end{tabular}
\label{tab:qg_parameters}
\end{table}

Mesoscale eddies emerge on a spatial scale determined by the deformation radius $r_d=\frac{g'}{f_0^2} \frac{H_1 H_2}{H}$ \cite{salmon1980baroclinic, vallis2017atmospheric}, denoted by the arrow in Figure \ref{fig:fig1}. We choose the resolution of the reference simulation ($256^2$) in order to accurately reproduce the spectral energy transfer. Coarse-resolution models do not resolve the deformation radius properly and fail to reproduce various statistical characteristics \cite{hallberg2013using, hewitt2020resolving}, including kinetic energy (KE), spectrum of KE and energy transfer. In this work, we aim to improve the simulation of turbulence in coarse models by incorporating a subgrid parameterization model, which compensates for the missing physics.

\subsection{Filtered equations}
 
 In this section, we derive the governing equations for the coarse model which follows the trajectory of the filtered and coarsegrained high-resolution simulation. 
 % the filtered fields, which should be numerically solved on a coarse mesh. 
 These equations contain a new term that describes the interaction with unresolved eddies, the term that is not available at the coarse resolution and needs to be parameterized. %An accurate parameterization of unresolved eddies can greatly improve the fidelity of the simulation of the turbulence on the coarse grid and thus gain computational efficiency compared to the approach when all mesoscale eddies are directly resolved on a fine grid.% the computational efficiency of the turbulence simulation. 
 %compared to the direct numerical simulation of all turbulent eddies.

We follow the Large eddy simulation (LES, \citeA{sagaut2006large}) approach to split the prognostic variables ($\phi$) into resolved ($\overline{\phi}$) and subgrid ($\phi'$) components by applying a spatial convolutional filter with kernel $G(\mathbf{y})$ such that
\begin{gather}
    \phi = \overline{\phi} + \phi', \\
    \overline{\phi} (\mathbf{x}) = \int G(\mathbf{y} - \mathbf{x}) \phi (\mathbf{y}) d \mathbf{y} ~~~ \mathrm{with} ~~~ 
    \int G(\mathbf{y}) d \mathbf{y} = 1. \label{eq:spatial_filter}
\end{gather}
We use two spectral filters from  \citeA{ross2022benchmarking}: one filter is a combination of a cut-off and a model filter ("Sharp"), the other is a combination of a cut-off and a Gaussian filter (we denote it as "Gaussian"). Precise definitions are provided in \ref{appendix:filters}.

Applying the filter $G(\mathbf{y})$ to the governing equations \eqref{eq:gov_eq_1}, \eqref{eq:gov_eq_2}, we obtain a set of governing equations for the filtered solution:
\begin{gather}
    \partial_t \overline{q}_m + \nabla \cdot (\overline{\mathbf{u}}_m \overline{q}_m) + \beta_m \partial_x \overline{\psi}_m + U_m \partial_x \overline{q}_m = - \delta_{m,2} r_{ek} \nabla^2 \overline{\psi}_m + S + ssd \circ \overline{q}_m, \label{eq:les_eq_1} \\
    \overline{q}_m = \nabla^2 \overline{\psi}_m + (-1)^m \frac{f_0^2}{g' H_m} (\overline{\psi}_1 - \overline{\psi}_2), ~ m \in \{1,2\}.
    \label{eq:les_eq_2}
\end{gather}
$S$ is the additional subgrid forcing produced by the unresolved eddies on the resolved scales,
\begin{equation}
    S = \nabla \cdot (\overline{\mathbf{u}} ~ \overline{q} - \overline{\mathbf{u} q}),\label{eq:subgrid_forcing}
\end{equation}
which needs to be parameterized.  
We will omit the index $m$ for the subgrid forcing and related variables to simplify notation. The dissipation term $ssd$ on a coarse grid in Eq. \eqref{eq:les_eq_1} is added a-posteriori to ensure the numerical stability of the simulations. In deriving Eq. \eqref{eq:les_eq_1}, we used commutativity between derivatives and spatial filtering, which holds for spectral numerical schemes and spectral filters \cite{ghosal1996analysis}. Both subgrid forcing and numerical advection scheme are formulated in flux form, so we include numerical approximation errors into the definition of subgrid forcing \cite{ghosal1996analysis, chow2003further, gullbrand2003effect}.

\subsection{Subgrid forcing dataset} \label{sec:dataset}
The solution to the governing equations \eqref{eq:gov_eq_1}, \eqref{eq:gov_eq_2} for the high-resolution model is denoted by $q$. The filtered quantities ($\overline{q}$) are defined on a coarse mesh.

\begin{figure}
\includegraphics[width=\textwidth]{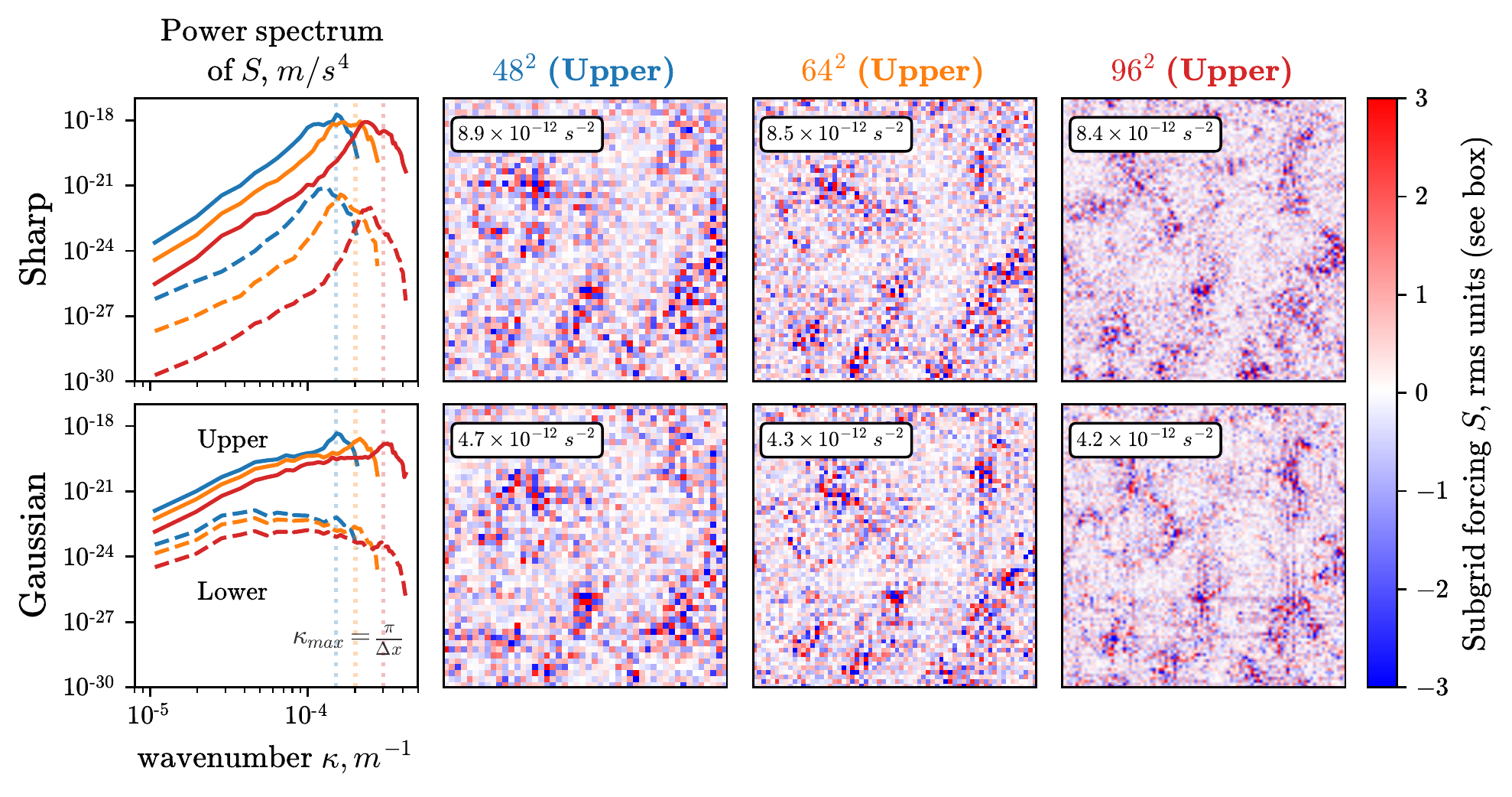}
\caption{
Subgrid forcing ($S$, Eq.~\eqref{eq:subgrid_forcing}) at different resolutions diagnosed using a Sharp filter (top row), or a Gaussian filter (bottom row). Left: power spectrum of $S$ for the upper fluid layer (solid lines) and lower fluid layer (dashed lines). Colors: $48^2$ (blue), $64^2$ (orange), $96^2$ (red). Vertical lines show grid cut-off for coarse mesh $\kappa_{max}=\pi/\Delta x$. Right: Snapshots of $S$ at three different resolutions for the upper layer.}
\label{fig:forcing_upper}
\end{figure}

The dataset to train ML subgrid parameterization models is obtained as follows. We integrate the governing equations in 
time for $10$ years at high resolution ($256^2$) with time step 1 hour
and save snapshots every 1000 hours, for a total of 86 snapshots. The training dataset consists of $250$ runs, each corresponding to a different random initial condition, for a total of $21500$ 
snapshots. The validation and testing datasets consist of $25$ runs each. 
For each  
coarse resolution ($48^2$, $64^2$, $96^2$), we compute 
a filtered solution represented on a coarse mesh ($\overline{q}$, $\overline{\mathbf{u}}$)
and subgrid forcing (Eq. \eqref{eq:subgrid_forcing}) using Sharp or Gaussian filters. The spectral content of the resulting subgrid forcing greatly depends on the scale selectivity of the filter, see Figure \ref{fig:forcing_upper}.

%\clearpage

\section{Data-driven stochastic subgrid models} \label{sec:subgrid_models}
In this section, we introduce a probabilistic approach for the prediction of subgrid forcing, which can be used to build data-driven stochastic parameterizations.

Conventional subgrid parameterizations %($\widetilde{S}$) 
establish a functional relationship % ($F$)
between the subgrid forcing ($S$, Eq. \eqref{eq:subgrid_forcing}) and the resolved flow ($\overline{q}$) in the form of $S \approx \widetilde{S}(\overline{q})$. 
% :
% \begin{equation}
%     \widetilde{S} = F(\overline{q}) \text{ with } S\approx \widetilde{S}. \label{eq:deterministic_mapping}
% \end{equation}
Such parameterizations are typically deterministic; they produce a single prediction for a given input. However, there is inherent uncertainty in the prediction of subgrid forcing, because many possible states of the subgrid eddies are consistent with a given resolved flow. Therefore, we propose to instead generate a \emph{probabilistic} prediction, by attempting to sample from the conditional distribution of the subgrid forcing given the coarse-grained flow  ($S \sim \rho(S|\overline{q})$).

\begin{figure}
\includegraphics[width=\textwidth]{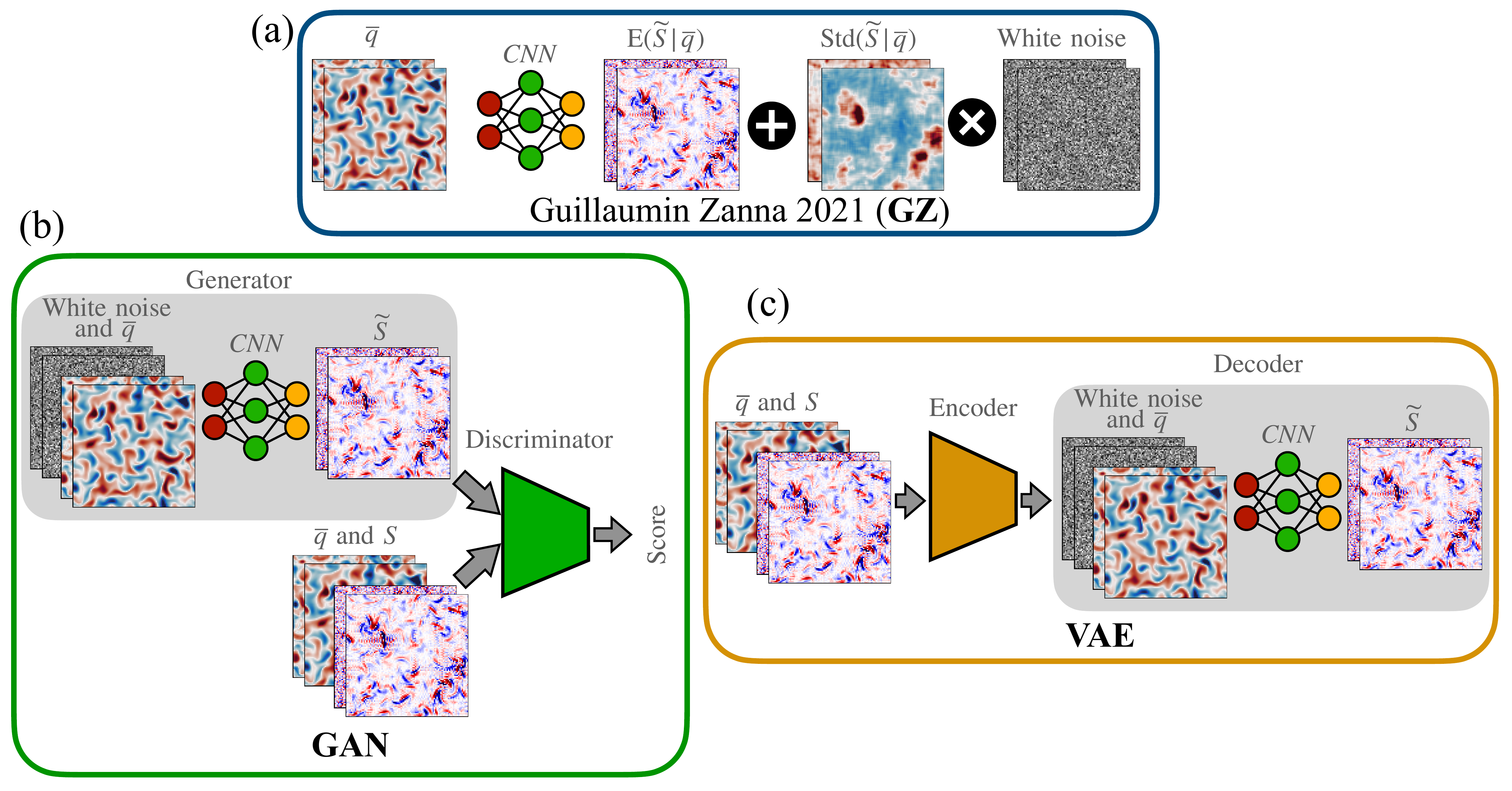}
\caption{Schematic of three stochastic subgrid models attempting to sample from conditional distribution $\rho(S|\overline{q})$: (a) GZ model, (b) GAN model and (c) VAE model. GZ model predicts uncorrelated stochastic residuals, but generative models (GAN and VAE) transform white noise using a mapping learned directly from data (gray-shaded box).
%can potentially generate correlated residuals because the white noise field is passed through CNN (gray-shaded box). 
Discriminator and Encoder are supplementary networks that allow training of the mapping but are not required for subgrid forcing prediction.
%Supplementary neural networks  (Discriminator and Encoder) are used only for training.
}
\label{fig:illustration}
\end{figure}

In order to generate a probabilistic prediction of the subgrid forcing, we propose to apply a generative ML framework, where samples from a desired distribution are obtained by transforming white noise using a mapping learned directly from data~\cite{kingma2013auto,goodfellow2014generative}. We design and compare three different approaches, depicted in Figure \ref{fig:illustration},  to learn this transformation: (a) A model based on \citeA{guillaumin2021stochastic}, which predicts the pointwise mean and pointwise standard deviation of the conditional distribution of the subgrid forcing. (b) A generative adversarial network (GAN), consisting of a \emph{generator} that generates subgrid-forcing samples by trying to fool a \emph{discriminator}, trained to distinguish between these samples and the true high-resolution data. (c) A variational autoencoder (VAE) consisting of an \emph{encoder}, which maps the input signal to a latent space, and a \emph{decoder}, which decodes the latent variables to produce subgrid-forcing samples. The remainder of this section provides a more detailed description of each approach.  
 
%All images are passed to neural networks in pairs, for upper and lower fluid layers. We use a similar convolutional neural network (CNN) for the generation of images in every stochastic model. In panel (a) we show a model analogous to \citeA{guillaumin2021stochastic} which predicts the pointwise mean and pointwise standard deviation of the conditional distribution. A stochastic part of the subgrid forcing is simulated by multiplication of the standard deviation field to a white noise field pointwise. This model does not account for the spatial correlation of stochastic residuals between grid points. To account for the spatial correlation in generative models GAN and VAE, we pass the white noise field to the input of CNN, see the gray-shaded box in panels (b) and (c). Training of generative models is accomplished with supplementary networks: Discriminator is used to train GAN and Encoder is used to train VAE. Below we derive the training algorithms for the presented models that can be skipped by the reader. We note that the main difference between models is in the loss functions, but the data generation process is similar: compare (Eq. \eqref{eq:GZ_parameterization}, GZ), (Eq. \eqref{eq:G_generator}, GAN) and (Eq. \eqref{eq:VAE_parameterization}, VAE).

\subsection{Guillaumin and Zanna model (GZ)} \label{sec:GZ_model}
\citeA{guillaumin2021stochastic} presented a probabilistic ML parameterization, where the mean and variance of the  subgrid forcing are estimated at every grid point using a neural network. %, where the mean and variance of the forcing based on an ,  using a model with Gaussian likelihood independent for every grid point and defined by the neural network predicting spatial fields of mean 
% \begin{equation}
%     \mathrm{E}(S|\overline{q})=\int S \rho(S|\overline{q}) dS
% \end{equation}
% and variance $\mathrm{Var}(S|\overline{q})=\mathrm{E}((S-\mathrm{E}(S|\overline{q}))^2|\overline{q})$ of the true conditional distirbution $\rho(S|\overline{q})$. 
The original formulation in \cite{guillaumin2021stochastic} minimizes an i.i.d. Gaussian likelihood cost function to optimize the parameters of the network. Here, we propose an alternative training procedure, which we have found to be more efficient. %which may give a suboptimal prediction of the conditional mean or an increased training time.
Following the approach of \citeA{adler2018deep}, we estimate the pointwise means and variances sequentially. 
% where $\mathrm{E}(S|\overline{q})$ and $\mathrm{Var}(S|\overline{q})$ are predicted by two regression problems. The method is based on the observation that optimization of the MSE loss gives a prediction of the conditional expectation \cite{bishop2006pattern}.

%We define two mappings $\widetilde{S}_{\theta}^{\mathrm{mean}}(\overline{q})$ and $\widetilde{S}_{\phi}^{\mathrm{var}}(\overline{q})$ which are intended to predict $\mathrm{E}(S|\overline{q})$ and $\mathrm{Var}(S|\overline{q})$, respectively, where $\overline{q}$ is the conditional variable. Both mappings are parameterized with the same CNN network (\ref{appendix:training_information}) with $\theta$, $\phi$ as the weights of neural networks.
%To ensure non-negative variance, we use a softplus activation function $\ln (1+e^{x})$ in the final layer of $\widetilde{S}_{\phi}^{\mathrm{var}}(\overline{q})$. 

First, we estimate the conditional mean at each grid point by minimizing the MSE loss function:
\begin{equation}
    \mathcal{L}_{\mathrm{MSE}}  = \frac{1}{2 n^2} || S - \widetilde{S}_{\theta}^{\mathrm{mean}}(\overline{q}) ||^2_2,
\end{equation}
where $\widetilde{S}_{\theta}^{\mathrm{mean}}(\overline{q})$ is the output of a neural network with parameters denoted by $\theta$, which receives $\overline{q}$ as an input. $S, \widetilde{S}_{\theta}^{\mathrm{mean}}, \overline{q} \in \mathbb{R}^{2 \times n \times n}$ are tensors representing two layers of fluid, each layer having $n \times n$ points. The norm in the cost function is the $\ell_2$ norm of the vectorized tensor, which for the vector of length $D$ is $|| \mathbf{x} ||_2=\sqrt{x_1^2+\cdots+x_D^2}$.
%We do not show in the loss function the averaging over training samples but it is present in the code. 
The loss function is minimized over a training set consisting of samples of the resolved flow $\overline{q}$ and the corresponding high-resolution forcing $S$ obtained as described in Section~\ref{sec:dataset}. 
Minimization of $\mathcal{L}_{\mathrm{MSE}}$ yields an optimal set of parameters $\theta^*$ and a corresponding model which we denote as $\widetilde{S}^{\mathrm{mean}}(\overline{q})$. 

Second, we estimate the conditional variance at each grid point, based on the residual of the conditional-mean estimate $r=S-\widetilde{S}^{\mathrm{mean}}(\overline{q})$. To this end, we minimize the cost function 
%Model predicting variance $\mathrm{Var}(S|\overline{q})$ is formulated as a model predicting $\mathrm{E}(r^2|\overline{q})$ that is the following loss is optimized:
\begin{equation}
    \mathcal{L}_{\mathrm{VAR}} = \frac{1}{2 n^2} || r^2 - \widetilde{S}_{\phi}^{\mathrm{var}}(\overline{q}) ||_2^2,\label{eq:loss_VAR}
\end{equation}
where $\widetilde{S}_{\phi}^{\mathrm{var}}(\overline{q})$ is the output of a neural network with parameters denoted by $\phi$, which receives $\overline{q}$ as an input. The final layer of the network is a softplus activation function $\ln (1+e^{x})$ to ensure that the variance estimates are nonnegative. 
The loss function is minimized over the training set fixing the residual $r$. The resulting model is denoted by $\widetilde{S}^{\mathrm{var}}(\overline{q})$. Additional training information is given in \ref{appendix:training_information}.
%Note that the resulting variance model depends on the quality of deterministic prediction $\widetilde{S}^{\mathrm{mean}}(\overline{q})$. As it will be discussed in Section \ref{sec:offline_analysis}, it may be beneficial, because  an inaccurate deterministic model should be accompanied by increased uncertainty. 

The conditional-mean and conditional-variance models are used to implement a stochastic parameterization with white noise, as follows (see Figure \ref{fig:illustration}(a)):
\begin{equation}
    \widetilde{S}(z,\overline{q}) = \widetilde{S}^{\mathrm{mean}}(\overline{q}) + (\widetilde{S}^{\mathrm{var}}(\overline{q}))^{1/2} \cdot z, \label{eq:GZ_parameterization}
\end{equation}
where $z \in \mathbb{R}^{2 \times n \times n}$ is sampled from a standard normal distribution. 

%\subsection{Deterministic model GZ(mean)}
%We consider the mean channel of GZ model as a deterministic model of subgrid forcing and denote it as "MSE"{}:
%\begin{equation}
%    \widetilde{S}(\overline{q}) = \widetilde{S}^{\mathrm{mean}}(\overline{q}).
%\end{equation}

\subsection{Generative adversarial network model (GAN)} \label{sec:GAN_model}
We propose to leverage the framework of generative adversarial networks (GANs) to build a probabilistic model \cite{goodfellow2014generative}, which generates samples from the distribution of possible subgrid forcings ($S$) at a given resolved flow ($\overline{q}$) denoted as $\rho(S|\overline{q})$, where both variables are considered as 3D fields, $S, \overline{q} \in \mathbb{R}^{2 \times n \times n}$. The mentioned distribution is defined implicitly by the dataset of pairs of $S$ and $\overline{q}$. The GAN framework consists of two networks, generator and discriminator, playing an adversarial game: the generator attempts to \emph{fool} the discriminator, which is trained to discriminate between the output of the generator and actual data sampled from a desired distribution. 
%As a result of the adversarial game, the generator learns to produce new samples from the desired distribution.
%We first give a definition of the generator which will be used as a stochastic parameterization and then present a way of its optimization which leads to the notion of discriminator. Finally, practical aspects of the optimization problem are considered.

Sampling from the conditional distribution is possible with the conditional GAN model (cGAN, \citeA{mirza2014conditional}), which informs both networks with the conditional variable. Specifically, the generator transforms the latent noise variable $z$ and PV field to the subgrid forcing, see  Figure \ref{fig:illustration}(b):
\begin{equation}
    \widetilde{S}=G(z, \overline{q}). \label{eq:generator_expression}
\end{equation}
The discriminator returns a score given a pair of subgrid forcing and PV field denoted as $D(S, \overline{q})$. 
%The output of the discriminator is a scalar scoring that subgrid forcing comes from the true distribution. 
There are many options to define the adversarial loss function \cite{lucic2018gans}. We leverage a popular approach of Wasserstein GAN (WGAN, \citeA{arjovsky2017wasserstein}) with the following optimization problem:
\begin{equation}
    \min_{G} \max_{D}% \in \mathrm{Lip}(S)} 
    \mathbb{E}
    \Big[ D(S, \overline{q})  - 
    D(G(z, \overline{q}), \overline{q}) \Big],
    \label{eq:GAN_loss_simple}
\end{equation}
where $\mathbb{E}$ is the mathematical expectation over the training samples. The discriminator $D$ is optimized to estimate the Wasserstein-1 distance between the distributions $\rho(S|\overline{q})$ and $\rho(\widetilde{S}|\overline{q})$,  while the generator learns the true distribution by minimizing this distance. 

Solving the optimization problem \eqref{eq:GAN_loss_simple} may lead to the mode collapse phenomenon when the generator ignores the latent variable $z$:
%because the dataset is unbalanced
for every coarse field $\overline{q}$ the model may produce a single fixed subgrid forcing \cite{isola2017image, ohayon2021high, mao2019mode, yang2019diversity}. To overcome mode collapse, we apply a technique proposed by \citeA{adler2018deep}: feeding multiple generator outputs $\widetilde{S}$ to the discriminator for a given input $\overline{q}$. 
Identical outputs are readily detected and penalized by the discriminator. We provide additional details in \ref{appendix:training_information}.

Once trained, the GAN generator \eqref{eq:generator_expression} can be used as a stochastic parameterization by sampling the latent  variable $z \in \mathbb{R}^{2 \times n \times n}$ from a standard normal distribution.

\subsection{Variational autoencoder model (VAE)}
As an alternative to the GAN framework, we propose leveraging the variational autoencoder (VAE, \citeA{kingma2013auto}) to sample from the conditional distribution $\rho(S|\overline{q})$. The VAE framework consists of two networks: the encoder and the decoder. The encoder produces a latent representation and the decoder reconstructs the subgrid forcing from this representation. A regularization term constrains the latent vector to be close to a simple distribution, chosen a priori. %This regularization allows to use the decoder as a generative ML model.

%In VAE architecture the decoder generates images of subgrid forcing and plays the same role as the generator in GAN. 
%The difference is that the decoder has an explicit (but simple) expression for the likelihood function denoted as $\rho_{\theta}(\widetilde{S}|z, \overline{q})$, see also Figure \ref{fig:illustration}(c). 
%Free parameters of the decoder and encoder will be denoted as $\theta$ and $\phi$, respectively. %and their optimal values as $\theta^*$ and $\phi^*$. 
%A special form of the decoder allows to define the conditional distribution of the subgrid forcing analytically \cite{kingma2013auto}:
% \begin{equation}
%     \rho_{\theta}(\widetilde{S}|\overline{q}) = \int \rho_{\theta}(\widetilde{S}|z, \overline{q}) \rho(z) dz, \label{eq:vae_total_probability}
% \end{equation}
% where $\rho(z)$ is a prior distribution of the latent variable. Following \citeA{sohn2015learning, doersch2016tutorial}, prior $\rho(z)$ does not depend on the conditional variable $\overline{q}$. Using an expression for total probability \eqref{eq:vae_total_probability}, estimation of the free parameters $\theta$ is possible by approximately maximizing the likelihood function $\rho_{\theta}(S|\overline{q})$, i.e. the probability density to observe the real subgrid forcing $S$ given the model $\rho_{\theta}$. Below we give a simplified derivation of the loss function of the conditional VAE. A complete explanation of VAE can be found in review papers \citeA{doersch2016tutorial, luo2022understanding}, 

The conditional VAE (cVAE) is obtained by feeding a conditional variable to the encoder and decoder \cite{sohn2015learning, doersch2016tutorial, zhang2016variational, mishra2018generative, pagnoni2018conditional}: 
%In our case, this conditional variable is the resolved flow $\overline{q}$. 
the decoder maps the latent noise and conditional variable $\overline{q}$ to the subgrid forcing, see Figure \ref{fig:illustration}(c):
\begin{equation}
    \widetilde{S} \sim \rho_{\theta}(\widetilde{S}|z, \overline{q}), \label{eq:decoder_expression}
\end{equation}
where free parameters are denoted as $\theta$ and we emphasize that the mapping is probabilistic. The probabilistic encoder with free parameters $\phi$ is denoted as $z \sim q_{\phi}(z|S, \overline{q})$. The encoder and decoder  are trained jointly to maximize the lower bound of the likelihood of observing the training sample (also known as evidence lower bound, ELBO):
\begin{eqnarray}
    \ln \rho_{\theta}(S|\overline{q}) \geq
     %\mathbb{E}_{q_{\phi}(z|S, \overline{q})} \ln \left( \frac{\rho_{\theta}(S|z, \overline{q})\rho(z)}{q_{\phi}(z|S, \overline{q})} \right) = \nonumber \\
     \underbrace{\mathbb{E}_{q_{\phi}(z|S, \overline{q})} \ln \rho_{\theta}(S|z, \overline{q})}_{\text{reconstruction}} - \underbrace{D_{\mathrm{KL}} \left(q_{\phi}(z|S, \overline{q}) || \rho(z) \right)}_{\text{regularization}} = - \mathcal{L}_{\mathrm{VAE}}, \label{eq:vae_loss}
\end{eqnarray}
where $D_{\mathrm{KL}}(p(x),q(x)) = \mathbb{E}_{p(x)}\ln\frac{p(x)}{q(x)}$ is Kullback–Leibler divergence, a measure of the difference between two distributions. 
%The loss function of VAE \eqref{eq:vae_loss} contains two terms. 
The reconstruction term encourages the encoder to seek an accurate latent representation of the subgrid forcing and encourages the decoder to assign a high probability to the training samples. The regularization term constrains the encoder to be close to the prior distribution $\rho(z)$. We parameterize all probabilities with Gaussian distributions and replace the mathematical expectation with one sample from the encoder (reparameterization trick, \citeA{kingma2013auto}). The resulting loss function is equivalent to a regularized MSE as explained in more detail in \ref{appendix:training_information}.

The mean channel of the Gaussian decoder \eqref{eq:decoder_expression} can be used as a stochastic parameterization by sampling the latent variable $z \in \mathbb{R}^{2 \times n \times n}$ from a standard normal distribution.

\section{Offline analysis of stochastic subgrid models} \label{sec:offline_analysis}
In this section, we perform an offline evaluation of the stochastic subgrid models described in Section \ref{sec:subgrid_models} using the dataset described in Section \ref{sec:dataset}. We show spatial maps and spectra of the predicted subgrid forcing. We propose metrics for the evaluation of the predicted subgrid forcing and stochastic residuals and compare them for a range of resolutions.
%We evaluate subgrid models with the We propose metrics for  Based on the observed properties of the subgrid models, we propose metrics for the offline analysis which are compared for a range of resolutions.

\begin{figure}
\includegraphics[width=\textwidth]{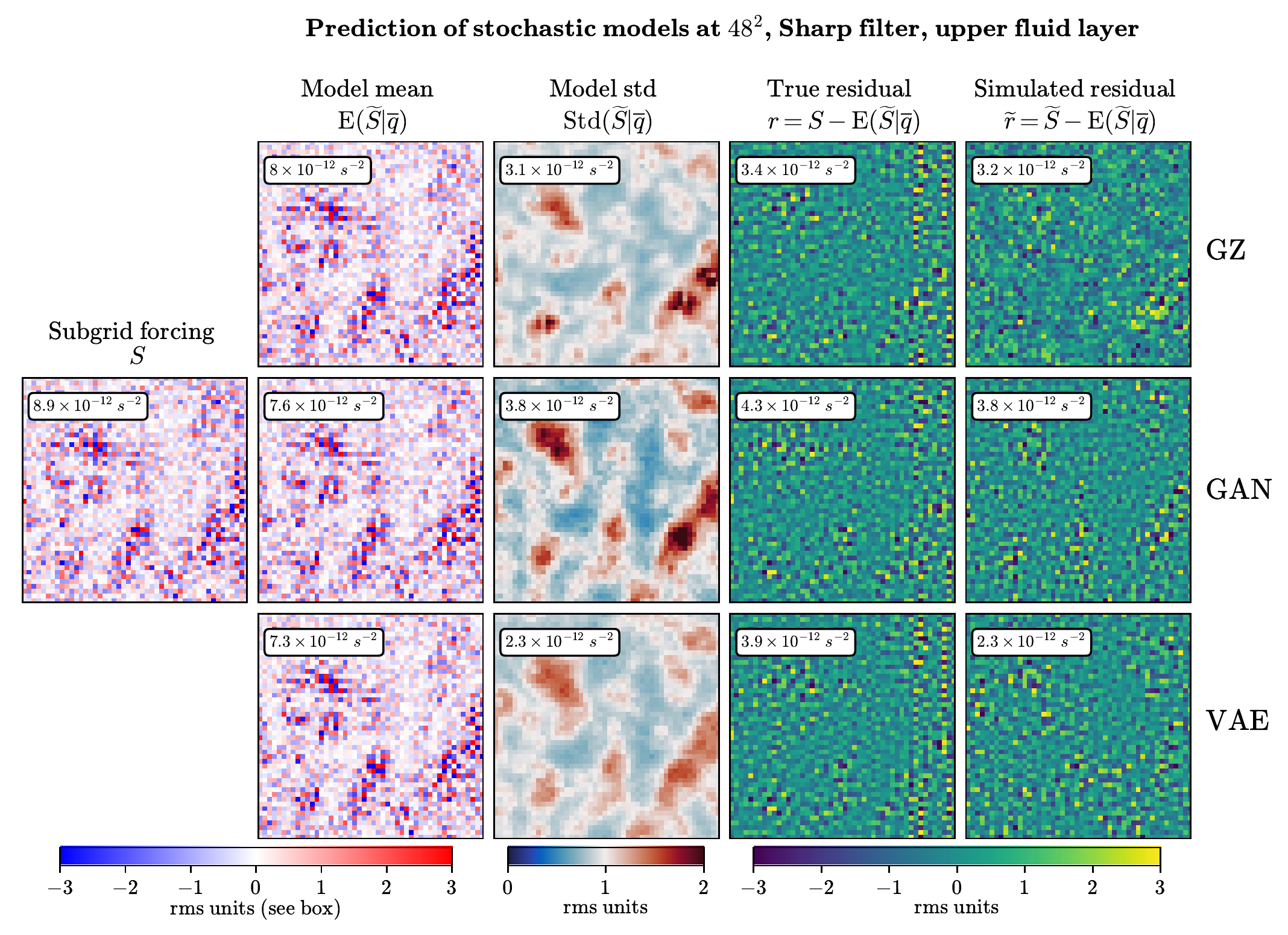}
\caption{Prediction of subgrid forcing ($S$) by stochastic models at a fixed conditional variable $\overline{q}$ on the testing dataset: GZ in upper row, GAN in middle row and VAE in lower row. "Model mean" is the deterministic prediction, and "Simulated residual" is the stochastic part of the model, which should be similar to "True residual". "Model std" is the standard deviation of the subgrid forcing prediction. The root mean square (rms) value of each field is shown in the box.
}
\label{fig:SGS_snapshot}
\end{figure}

For every combination of filter (Sharp and Gaussian) and resolution of coarse mesh ($48^2$, $64^2$, $96^2$), we train three machine learning models: GZ, GAN and VAE, for a total of 18 different models of subgrid forcing. 
The baseline deterministic subgrid model is trained with the MSE loss function and it is referred to as "MSE"{} (we simply take the mean channel of GZ model). Following \citeA{kochkov2021machine}, every model was trained  five times with different random seeds. Three  training instances failed and were excluded from the subsequent analysis: 2 realizations of VAE models at resolution $64^2$ experienced the posterior collapse problem (zero spread, \citeA{dai2020usual}), and one realization of GZ model at resolution $48^2$ had a large generalization error.
In this section, we show results for Sharp filter, and similar plots for Gaussian filter are shown in \ref{appendix:offline}.

\subsection{Analysis of stochastic predictions}

%An individual prediction of the subgrid forcing by a stochastic parameterization requires to sample a white noise field ($z$) and to provide a conditional variable ($\overline{q}$). The final expressions are similar:  compare (Eq. (18), GZ), (Eq. (20), GAN) and (Eq. (39), VAE). Conditional mean and conditional variance are directly accessible for the GZ model, and for GAN and VAE models they are estimated using 1000 samples, see \citeA{adler2018deep}.
In this section, we compare stochastic predictions of subgrid forcing to the true subgrid forcing. We suggest to split the stochastic prediction into the deterministic part and the stochastic residual. We define the deterministic part as a mean prediction of subgrid forcing at a fixed resolved field $\overline{q}$  -- it is conditional mean denoted as $\mathrm{E}(\widetilde{S}|\overline{q})$.
The deterministic part of the GZ model is given by the mean channel $\widetilde{S}^{\mathrm{mean}}(\overline{q})$, and for GAN and VAE models we estimate it by sampling $1000$ realizations of the latent vector \cite{adler2018deep}. The predicted stochastic residuals ($\widetilde{r} = \widetilde{S} - \mathrm{E}(\widetilde{S}|\overline{q})$) should be compared to the true residuals ($r = S-\mathrm{E}(\widetilde{S}|\overline{q})$), and for an accurate stochastic model they should be statistically similar, see \citeA{wilks2005effects, arnold2013stochastic, agarwal2021comparison, berner2009spectral, mana2014toward, shutts2007convective, guillaumin2021stochastic, gagne2020machine}.

In Figure \ref{fig:SGS_snapshot} we show predictions of the stochastic subgrid models. The deterministic part ($\mathrm{E}(\widetilde{S}|\overline{q})$) is similar for three stochastic models. The predicted stochastic residual (rightmost column) for GZ model looks like uncorrelated spatial white noise in contrast to the true residual. The predicted residuals for the other two models (GAN and VAE) are more visually similar to the true one. The pointwise standard deviation is a measure of the local uncertainty in the deterministic prediction and can be related to the second moment of residuals as:
$$
\mathrm{Std}(\widetilde{S}|\overline{q}) = \sqrt{\mathrm{E}\left(\left(\widetilde{S}-\mathrm{E}(\widetilde{S}|\overline{q})\right)^2\Big|\overline{q}\right)} = \sqrt{\mathrm{E}(\widetilde{r}^2|\overline{q})}.
$$
It is directly accessible for the GZ model, and for GAN and VAE models it can be estimated similarly to the conditional mean. The standard deviation fields have similar spatial structure for all three stochastic models.

\begin{figure}[h!]
\centering
\includegraphics[width=1\textwidth]{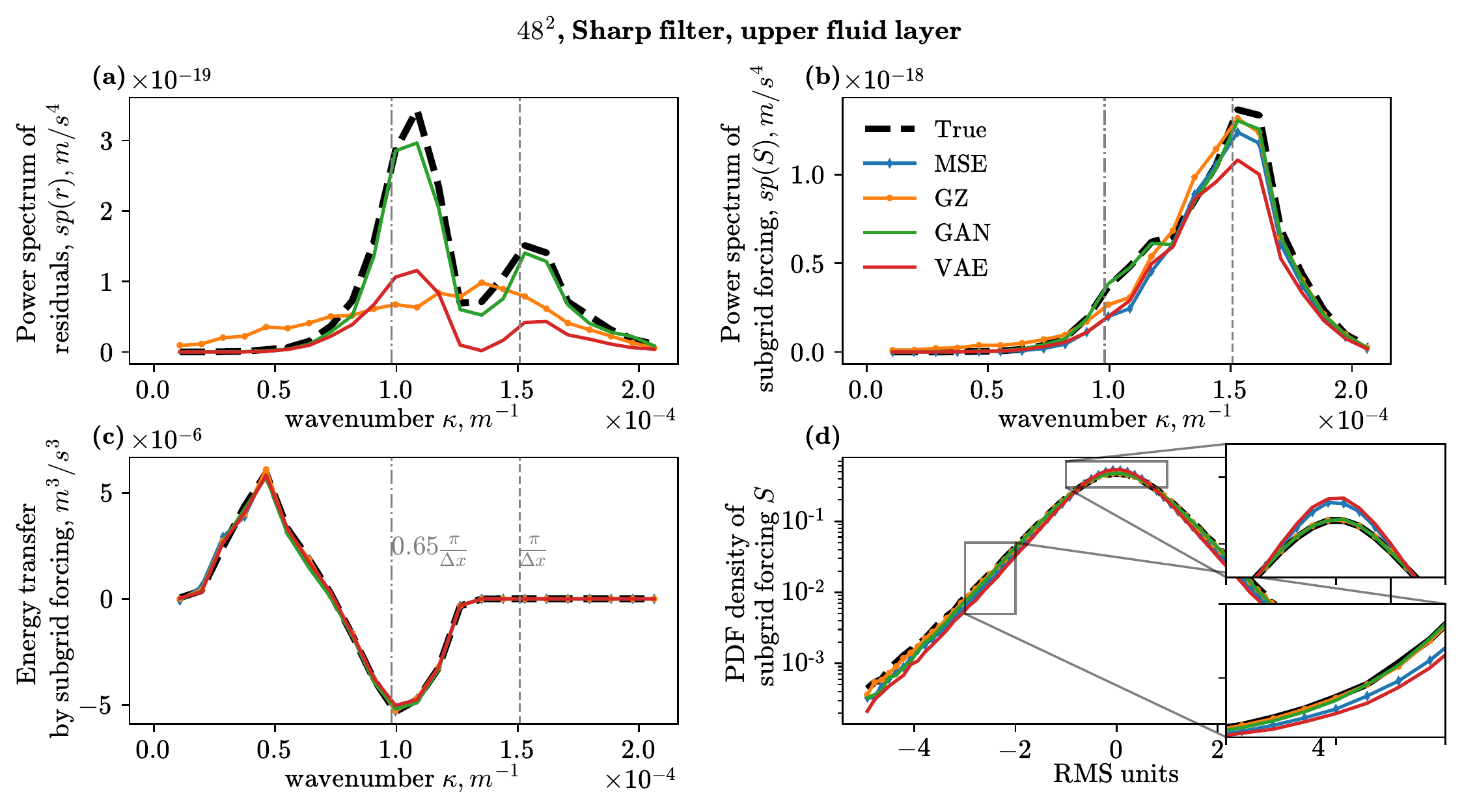}
\caption{Offline analysis of stochastic subgrid models (GZ, GAN, VAE): (a) power spectrum of stochastic residuals and (b) subgrid forcing; (c) energy transfer (see Eq. \eqref{eq:energy_balance}) and (d) marginal PDF of subgrid forcing.
MSE is the deterministic subgrid model given by the mean channel of GZ model. }
\label{fig:spectra_offline}
\end{figure}

We use the spatial power spectrum to analyze the spatial correlation.
 In Figure \ref{fig:spectra_offline}(a) we show the power spectrum of stochastic residuals. The true residuals are concentrated near the grid cut-off (Nyquist frequency, $\kappa_{max}=\pi/\Delta x$) and near the spatial frequency of $ssd$ filter ($\kappa=0.65\pi/\Delta x$). The GZ model does not reproduce the two-hill shape of the power spectrum of residuals. The GAN model accurately reproduces the power spectrum of residuals and improves the power spectrum of subgrid forcing (Figure \ref{fig:spectra_offline}(b)) compared to the deterministic and stochastic baselines (MSE, GZ). Note that accurate prediction of the power spectrum of subgrid forcing is a challenging task for deterministic models \cite{guan2022learning} because optimization of the mean squared error leads to the loss of details in small scales \cite{isola2017image}. The VAE model predicts the correct shape of the spectrum of residuals, but the total variance of residuals is underestimated. The power spectrum of subgrid forcing for the VAE model is also lower compared to the other models. We explain it by the well-known issue of VAE architecture to predict oversmoothed images \cite{takida2022preventing}.
 
An important property of subgrid forcing in QG turbulence is an ability to energize turbulence on a coarse grid, i.e. kinetic energy backscatter \cite{jansen2014parameterizing}.
%\cite{frederiksen1997eddy, meneveau2000scale, berner2009spectral, thuburn2014cascades, jansen2014parameterizing, zanna2017scale, juricke2020ocean}. 
There are two popular approaches to simulate backscatter: stochastic residuals near the grid scale \cite{leslie1979application, chasnov1991simulation, schumann1995stochastic, frederiksen1997eddy, grooms2015numerical} and mean energy injection in large scales \cite{kraichnan1976eddy, frederiksen2003effects, graham2013framework, thuburn2014cascades, jansen2014parameterizing, juricke2020ocean}. These two types of backscatter result from physical processes of a very different nature: stochastic backscatter simulates the loss of information about unresolved degrees of freedom but energy injection in large scales compensates for the unresolved inverse energy cascade. All the stochastic models are accurate in predicting large-scale energy injection (Figure \ref{fig:spectra_offline}(c)), and GAN model is the best in predicting stochastic residuals near the grid scale. 

Marginal PDF of subgrid forcing is often used to evaluate subgrid models \cite{pawar2020priori, maulik2017neural}. Both GZ and GAN models improve this PDF in the high-probability region and in the tails compared to the baseline MSE model, see Figure \ref{fig:spectra_offline}(d). The VAE model is similar to the baseline MSE in this characteristic.

%The simulation of stochastic residuals does not influence the average energy transfer of subgrid forcing, see Figure \ref{fig:spectra_offline}(c). We explain it as follows: the average energy exchange with subgrid scales is a linear function of subgrid forcing $\mathrm{E}(\overline{\psi} S)$, and thus uncorrelated residuals do not change it: $\mathrm{E} (\overline{\psi} S)= \mathrm{E} (\overline{\psi} \mathrm{E}(S|\overline{q}))$, see also \citeA{moser2021statistical}. \citeA{guan2022learning} show that various deterministic models are able to capture the mean energy transfer of subgrid forcing, but its power spectrum is underestimated. Here we show that the generative stochastic model GAN can capture both characteristics. An inability to accurately reproduce the power spectrum and PDF of subgrid forcing by a deterministic model (such as MSE) can be explained by a widely recognized problem in Machine Learning known as "regression to the mean"{} \cite{isola2017image}. That is, the optimal deterministic model -- conditional mean $\mathrm{E}(S|\overline{q})$ -- predicts blurred images due to averaging operation inherent to expectation.

\subsection{Quantitative offline analysis and metrics}
\begin{table}[h!]
\begin{tabular}{l|cccc}
Metric & $\mathcal{L}_{\mathrm{rmse}}$                           & $\mathcal{L}_{\mathrm{S}}$              & $\mathcal{L}_{\mathrm{r}}$              & $\sigma^2_{\mathrm{spread}}$           \\
\hline
Expression&$\frac{||S-\mathrm{E}(\widetilde{S}|\overline{q})||_2}{||S||_2}$ & $\frac{||sp(S)-sp(\widetilde{S})||_2}{||sp(S)||_2}$ & $\frac{||sp(r)-sp(\widetilde{r})||_2}{||sp(r)||_2}$ & $\frac{||\widetilde{r}||^2_2}{||r||^2_2}$ \\
Optimal value & 0 & 0& 0& 1 \\
Unparameterized model & 1 & 1 & 1& 0 \\
Quality of & deterministic part & full forcing & residuals & residuals 
\end{tabular}
\caption{Metrics for offline analysis of stochastic subgrid model $\widetilde{S}$: RMSE of deterministic part ($\mathcal{L}_{\mathrm{rmse}}$), RMSE in spectrum of the full subgrid forcing ($\mathcal{L}_{\mathrm{S}}$), RMSE in spectrum of stochastic residuals ($\mathcal{L}_{\mathrm{r}}$) and spread of the samples of conditional distribution ($\sigma^2_{\mathrm{spread}}$). We denote computaiton of isotropic power spectrum as $sp(\cdot)$.}
\label{tab:offline_metrics}
\end{table}

\begin{figure} [h!]
\centering
\includegraphics[width=1\textwidth]{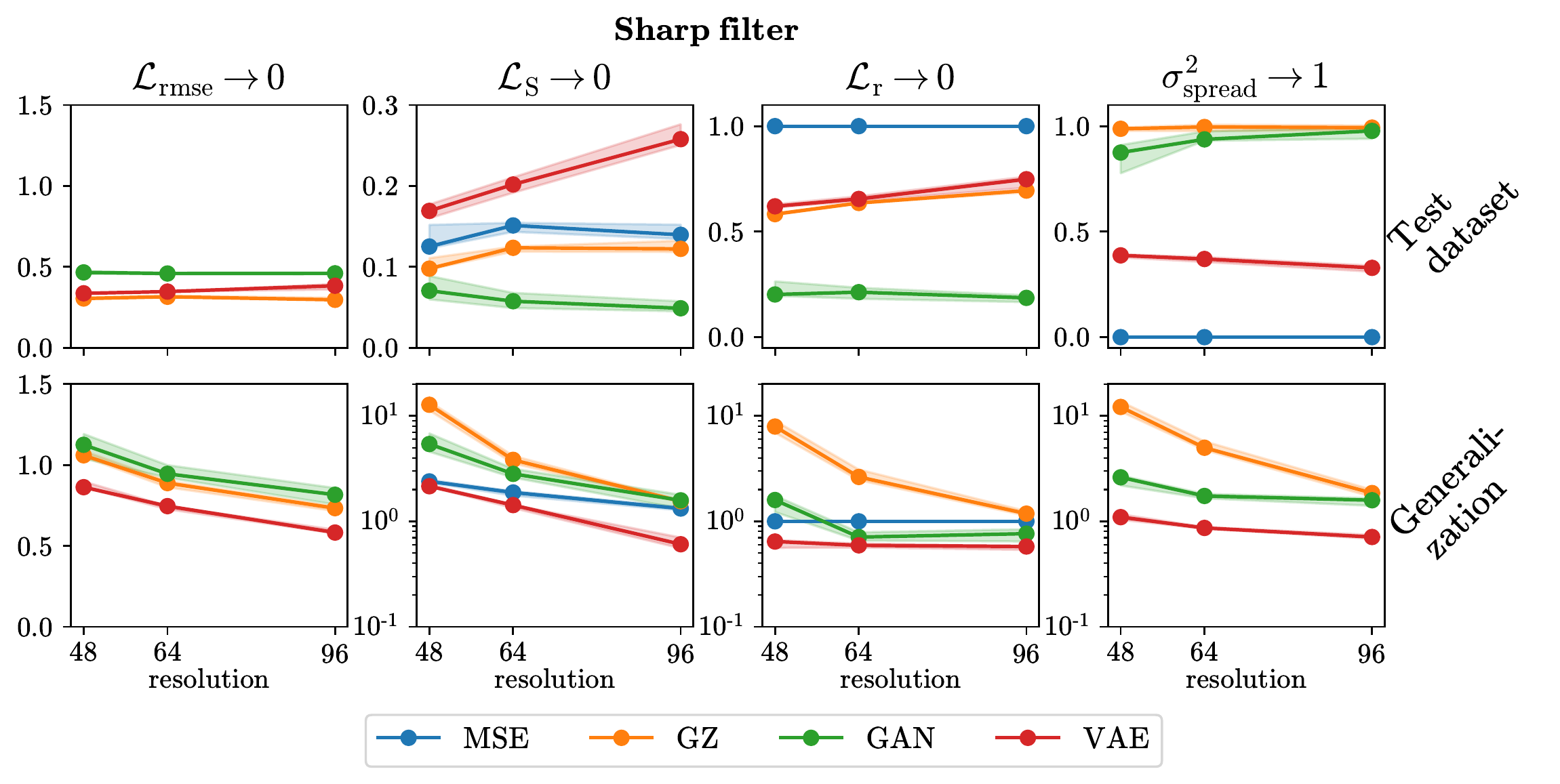}
\caption{Offline metrics from Table \ref{tab:offline_metrics} on testing dataset in the upper row and generalization to configuration with jets \cite{ross2022benchmarking} in the lower row. Optimal values are given with arrows. Each model is trained 5 times with different random seeds. The shading area shows min-max values among training realizations, markers show median value.}
\label{fig:offline_metrics}
\end{figure}

Above we presented a qualitative analysis of the stochastic subgrid models, and here we propose metrics for their quantitative evaluation.
We consider three classes of metrics, which demonstrate: the quality of the subgrid forcing, its deterministic part and stochastic residuals, see Table \ref{tab:offline_metrics}. We include spectral metrics for the subgrid forcing and residuals ($\mathcal{L}_{\mathrm{S}}$  and $\mathcal{L}_{\mathrm{r}}$) in order to evaluate to what extent the models capture the corresponding spatial structure.

In Figure \ref{fig:offline_metrics} we report the evaluation of the offline metrics for the different models. The upper row provides metrics on the test dataset with the same turbulence regime as the training set. We observe that the generative models (GAN and VAE) have slightly greater deterministic error ($\mathcal{L}_{\mathrm{rmse}}$) compared to the model optimizing this metric directly (GZ and MSE). The GAN and GZ models correctly predict spread of stochastic residuals $\sigma^2_{\mathrm{spread}} \approx 1$, but the VAE model underestimates spread $\sigma^2_{\mathrm{spread}} \approx 0.35$. The GAN model clearly outperforms the rest in predicting the spectra of the subgrid forcing $\mathcal{L}_{\mathrm{S}}$ and the residuals $\mathcal{L}_{\mathrm{r}}$. The VAE model on the contrary has high errors $\mathcal{L}_{\mathrm{S}}$ and $\mathcal{L}_{\mathrm{r}}$ because it predicts oversmoothed samples with reduced diversity.

\begin{figure} [h!]
\centering
\includegraphics[width=0.6\textwidth]{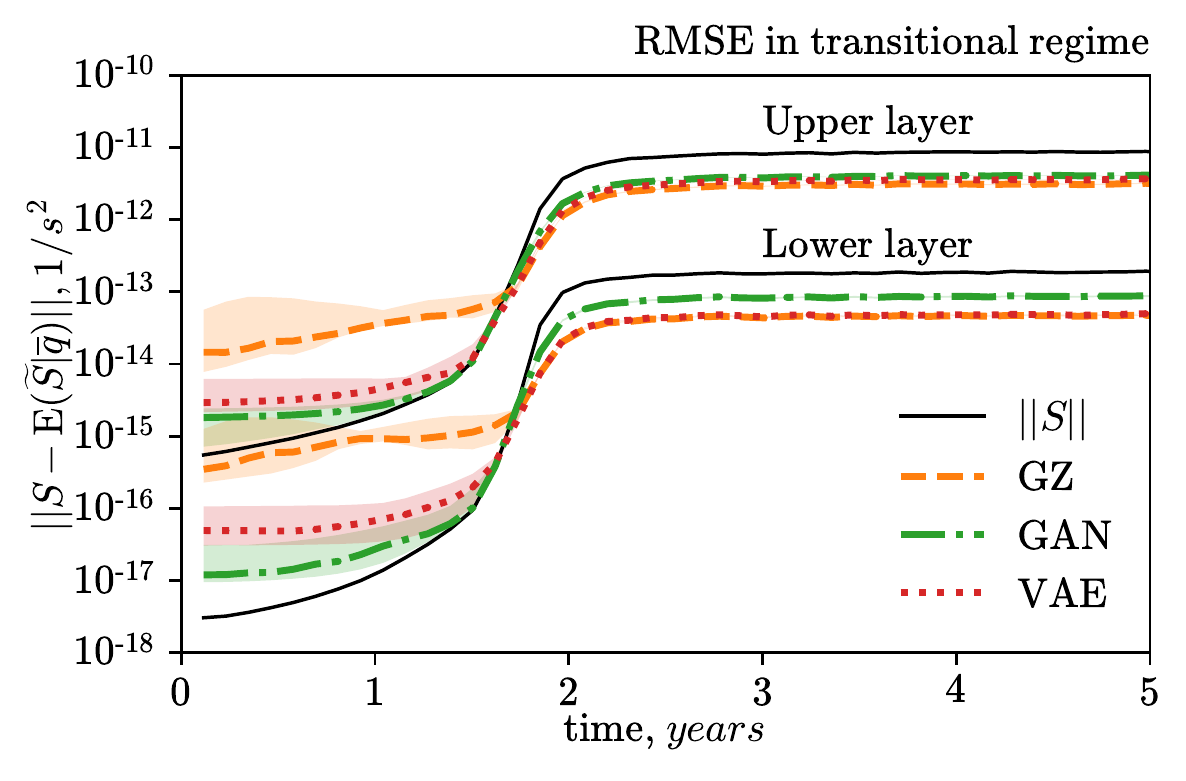}
\caption{Offline analysis of RMSE of deterministic part in transitional regime ($t< 2$ years) on test dataset. Norm $||\cdot ||$ is given per one grid point. Shading corresponds to different training realizations of the same model. Sharp filter, resolution $48^2$. Each line is given twice: for lower and upper fluid layer.}
\label{fig:offline_transitional}
\end{figure}

In the lower row of Figure \ref{fig:offline_metrics}, we evaluate the generalization ability of the models by computing the offline metrics on  dataset corresponding to a different turbulence regime, where flow is dominated by meandering jets, and which is therefore systematically different from the training data (see \citeA{ross2022benchmarking} for description). GZ model considerably overestimates the spread of the residuals ($2<\sigma^2_{\mathrm{spread}}<10$), and it deteriorates the spectral metrics ($\mathcal{L}_{\mathrm{S}}$  and $\mathcal{L}_{\mathrm{r}}$).
%Large spread considerably deteriorates the quality of individual samples as indicated by large error $\mathcal{L}_{\mathrm{S}}$ for GZ model compared to its deterministic part GZ(mean). 
Although the VAE model had various suboptimal metrics  on the eddy turbulence configuration, it demonstrates the best generalization capabilities to the jet configuration: VAE model has reasonable spread $\sigma^2_{\mathrm{spread}} \approx 0.8$, and outperforms other models in the error of the deterministic prediction $\mathcal{L}_{\mathrm{rmse}}$, the quality of the subgrid forcing $\mathcal{L}_{\mathrm{S}}$ and residuals $\mathcal{L}_{\mathrm{r}}$. The GAN model generalizes better than GZ for most of the metrics, without reaching the performance of the VAE model. We observe similar results for the Gaussian filter, see \ref{appendix:offline}.

During the first few years of simulation, QG model undergoes a transition from a laminar to a turbulent flow regime. Generalization to the transitional regime is a difficult test for subgrid models \cite{frezat2021physical} because the subgrid forcing is a few orders of magnitude smaller compared to the developed turbulence regime. Although we include the transitional regime in the training set, the relative importance of these samples is small due to their small norm. As a result, all subgrid models have large errors compared to the norm of the subgrid forcing during the first few years of simulation ($t < 2$ years), see Figure \ref{fig:offline_transitional}. 
The generative models (GAN and VAE) demonstrate the best performance in the transitional regime: error is one order of magnitude smaller compared to GZ. In the next section, we show that generative models are also superior to the baseline in the online simulation of transitional flow.

\section{Online simulations with subgrid models} \label{sec:online_analysis}
In the previous section, we demonstrated the encouraging ability of generative models GAN and VAE to simulate various statistical characteristics of subgrid forcing. 
%We expect that accurate modeling of the stochastic part of subgrid forcing may improve statistical properties of long "climate"{} simulations \cite{palmer2000predicting}. 
In this section, we evaluate the performance of trained subgrid models in online simulations. In more detail, we use the output of the subgrid model $\widetilde{S}$ to replace the true subgrid forcing $S$ in the governing equation for the coarsegrained dynamics \eqref{eq:les_eq_1}, and perform numerical time integration. Our goal is to study how the subgrid parameterizations impact the dynamics of mesoscale eddies in a statistical equilibrium regime.

Our online experiments are summarized in Table \ref{tab:qg_parameters}. Compared to the generation of the training data, we run experiments for twice as long (20 years). Before passing the subgrid forcing prediction into governing equation we subtract the spatial mean in each fluid layer to ensure the conservation of PV. Recall that we train 5 different models (differing only in the initialization of the weights) for every combination of resolution, filter and type of subgrid model. Each of these models is evaluated in an ensemble of 10 online runs, with different random initial conditions. The total number of runs is approximately $1200$. The statistical characteristics of the turbulence are averaged over the 10 ensemble members (and the last 15 years if applicable). We provide the confidence bounds for every averaged statistic defined by the minimum, maximum and median values over 5 realizations of the training algorithm.
%An approximate number of individual online runs is $1200$.
%\begin{gather*}
%    \text{3 resolutions} \times \text{ 2 filters } \times \\ \text{ 4 ML-subgrid models } \times \text{ 5 realizations of training } \times \text{ 10 ensemble runs } = 1200.
%\end{gather*}
%The small-scale dissipation term $ssd$ is retained and preserves the numerical stability of parameterized QG models. 

%Note that these simulations became unstable because ML parameterizations simulate large-scale kinetic energy backscatter and may accidentally over-energize the flow in an uncontrolled way compared to the energetically-consistent parameterizations \cite{jansen2014parameterizing}.
%because we do not control the amplitude of the parameterization as it is usually done in energetically-consistent physical parameterizations of backscatter \cite{jansen2014parameterizing}.

Among mentioned experiments, there were a few unstable simulations at resolution $96^2$: one run of the VAE model for Sharp filter, and 3 GAN models out of 5 training realizations for Gaussian filter. We exclude mentioned experiments from the analysis. In these experiments, an eddy emerges which is constantly amplified by the parameterization, and in a spectral space it corresponds to the overestimated energy injection on the largest scale. This effect is possible because we do not control the amplitude of the parameterization as it is usually done in energetically-consistent physical parameterizations of backscatter \cite{jansen2014parameterizing}.

\subsection{Metrics for online analysis}
We compare the solution of the coarse parameterized model to the filtered and coarse-grained fields of the high-resolution model similarly to \citeA{nadiga2007instability, beck2019deep, frezat2022posteriori, guan2022stable, guan2022learning}.

%The solution of the governing equations for the coarse system Eq. \eqref{eq:les_eq_1}, \eqref{eq:les_eq_2} would exactly reproduce the coarsegrained and filtered turbulence fields $\overline{q}$ for an ideal subgrid model that perfectly estimates the subgrid forcing \cite{nadiga2007instability, beck2019deep}. 
%Similarly to \citeA{frezat2022posteriori, guan2022stable, guan2022learning},
%we evaluate the model by checking to what extent it is able to reproduce the statistical characteristics of filtered and coarsegrained solution.% 
%we assume that a practical subgrid model which cannot exactly predict subgrid forcing still should be able to reproduce statistical characteristics of filtered and coarsegrained solution.

Following \citeA{ross2022benchmarking}, we consider an error in PDFs of the turbulence fields. Define the Wasserstein distance between distributions as $\mathcal{W}_1(F_1, F_2) = \int |F_1(\xi) - F_2(\xi)| d\xi$, where $F_1$ and $F_2$ are cumulative distribution functions (CDF) of some variable $\xi$. In computing CDF, we aggregate spatial directions, 15 years of simulation, and 10 ensemble members. We consider 5 variables in place of $\xi$: potential vorticity ($q_m$), velocity ($u_m$ and $v_m$), kinetic energy ($\frac{1}{2}|\mathbf{u}_m|^2$) and relative enstrophy ($\frac{1}{2}|\mathrm{curl}(\mathbf{u}_m)|^2$), and each fluid layer is accounted independently. The online distributional metric between the coarse-grid model  ($F_{\mathrm{model}}$) and the filtered and coarse-grained high-resolution simulation ($F_{\overline{\mathrm{hires}}}$) is given by the average of normalized errors:
\begin{gather}
    \mathcal{W}(\mathrm{model}, \overline{\mathrm{hires}}) = \frac{1}{10} \sum_{m=1}^2 \sum_{\xi \in \mathrm{Vars}_m } \frac{\mathcal{W}_1 \left(F_{\mathrm{model}}(\xi), F_{\overline{\mathrm{hires}}}(\xi) \right)}{ \sqrt{ \int \xi^2 d F_{\overline{\mathrm{hires}}}}},
    \label{eq:distribution_metric}
\end{gather}
where $\mathrm{Vars}_m = \{q_m, u_m, v_m, \frac{1}{2}|\mathbf{u}_m|^2, \frac{1}{2}|\mathrm{curl}(\mathbf{u}_m)|^2\}$ and the normalization constant is the square root of the uncentered second moment.

An additional metric based on spectral characteristics is reported in \ref{appendix:online}.

\subsection{Sensitivity to the correlation time of latent variable} \label{sec:time_sampling}
%\subsection{Time sampling method}
\begin{figure} [h!]
\centering
\includegraphics[width=0.5\textwidth]{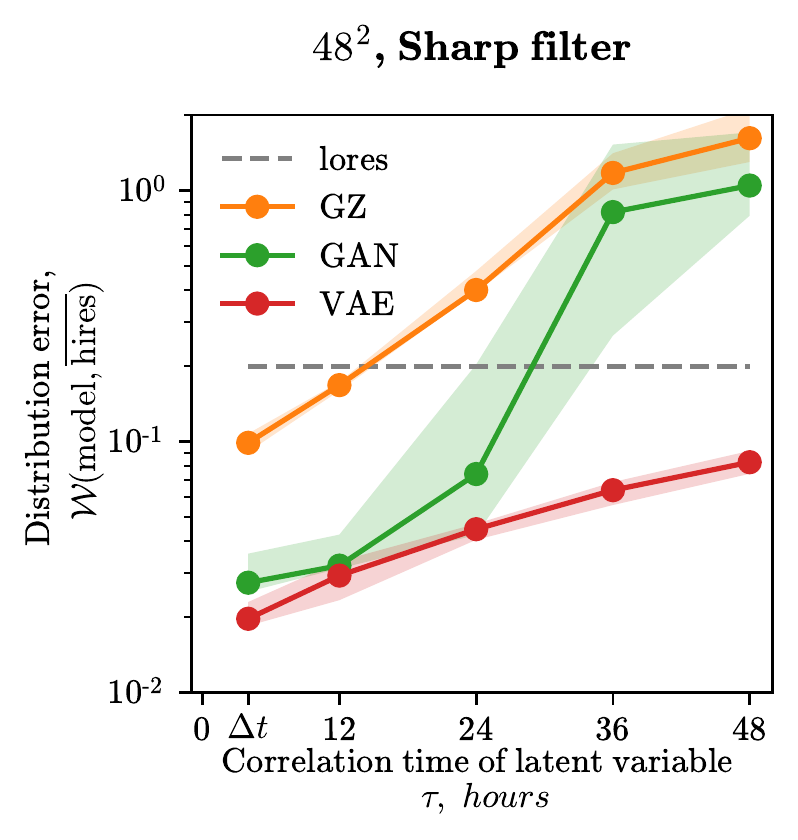}
\caption{Online distributional metric \eqref{eq:distribution_metric} as a function of correlation time $\tau$ of latent variable $z$ for three stochastic subgrid models (GZ, GAN and VAE). lores is model on a coarse grid without parameterization. Shading area shows min-max values
among training realizations, and markers show median value. Time step of the numerical integration $\Delta t$ is 4 hours.}
\label{fig:correlation_time}
\end{figure}

In order to leverage the proposed subgrid-forcing models in a stochastic parameterization, we sample the latent variable $z$ independently at every time step (discrete white noise) similar to \citeA{zanna2017scale} and \citeA{guillaumin2021stochastic}. 

Following \cite{gagne2020machine}, we also tested the sensitivity of the online simulation results to the correlation time of the latent variable. The time correlation is introduced with the autoregressive model of order one (AR1), which has covariance function $\mathrm{E} (z(t) z(t+n\Delta t)) = (1 - \Delta t / \tau)^n$ \cite{schumann1995stochastic}, where $n$ denotes the number of time layers between two time moments,  $\tau \geq \Delta t$ is correlation time, and at $\tau=\Delta t$ we restore the discrete white noise process. The online distributional metric \eqref{eq:distribution_metric} as a function of correlation time is reported in Figure \ref{fig:correlation_time}. The optimal online metric corresponds to $\tau=\Delta t$, which justifies our method of sampling (white noise).

\subsection{Results}
\begin{figure} [h!]
\centering
\includegraphics[width=1\textwidth]{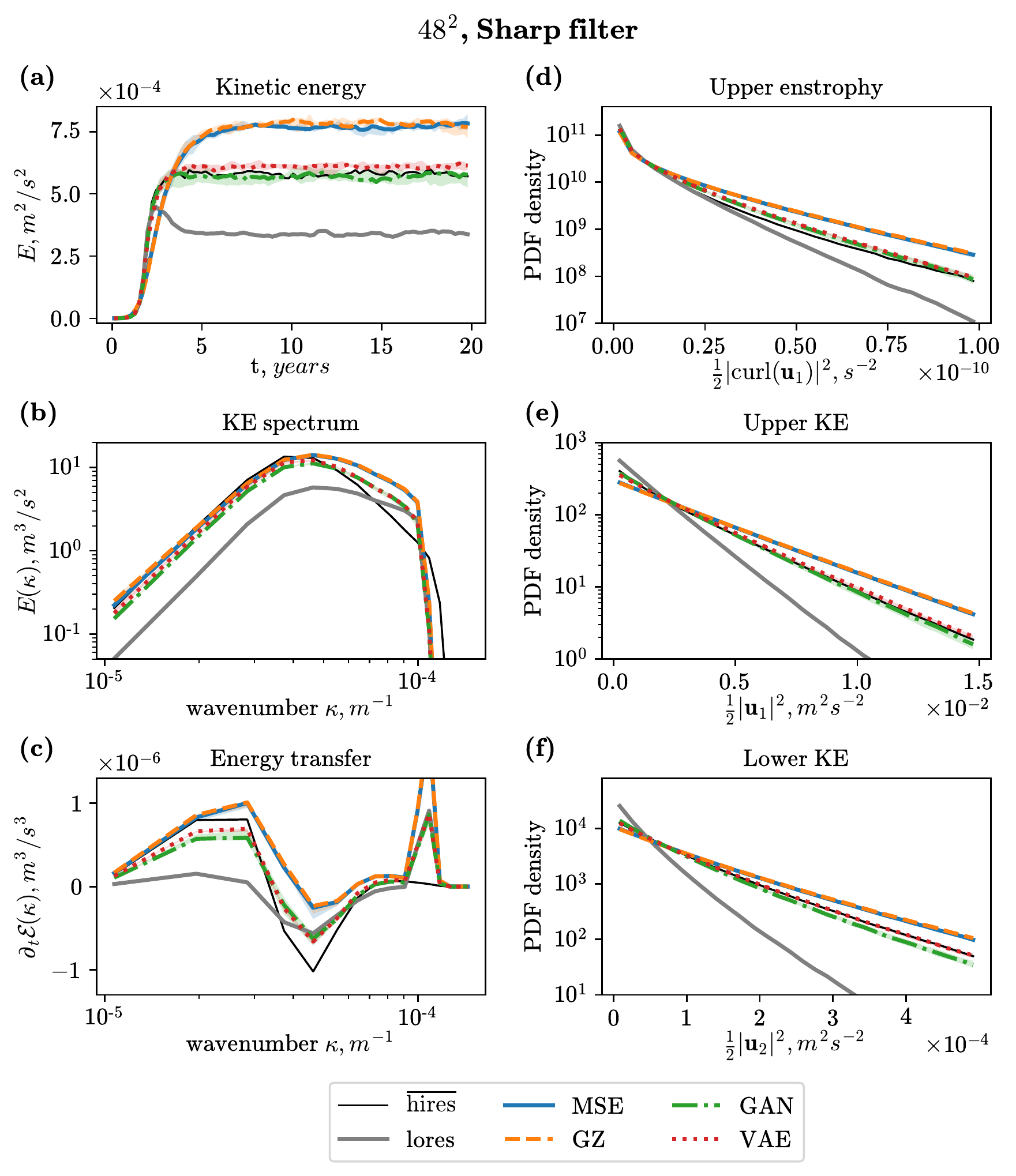}
\caption{Online simulations with parameterized models (MSE, GZ, GAN and VAE). $\mathrm{lores}$ is a coarse unparameterized model. $\overline{\mathrm{hires}}$ is filtered and coarsegrained high-resolution model. MSE is a deterministic subgrid model given by the mean channel of the GZ model. Energy transfer on panel (c) gives a sum of contributions from the resolved advection and subgrid model (see Eq. \eqref{eq:energy_balance}). Shading area shows min-max values among training realizations, and lines show median value. The time step $\Delta t$ is 2 hours.}
\label{fig:online_48}
\end{figure}

In Figure \ref{fig:online_48} we show online simulations with subgrid models at the coarsest resolution. The unparameterized model ("lores"{}) has underestimated kinetic energy (a) and underestimated KE spectrum in large scales (b). This is due to the poor representation of the inverse energy cascade on the coarse grid (c). The deterministic subgrid model (MSE) improves inverse energy cascade and KE spectrum in large scales, but small eddies near the grid scale are energized too much, see KE spectrum in small scales, KE level and tails of PDFs. The GZ model does not prevent overamplification of the small eddies. 
%We expect that the stochastic part of the GZ model cannot change significantly the dynamics of the small eddies, because uncorrelated spatial noise in the potential vorticity equation injects energy on the domain scale  (see Figure 7 in \citeA{jansen2014parameterizing}). 
In contrast, the generative stochastic models (GAN and VAE) improve the simulation of the small eddies: see spectral characteristics in small scales, tails of PDFs and kinetic energy. 
%The improvement in the dynamics of the small eddies by the generative models (GAN and VAE) may be connected to an accurate simulation of stochastic residuals near the grid scale (stochastic backscatter) or to better properties of the deterministic part of these subgrid models. \textcolor{red}{In additional experiments with the deterministic part of the generative models (approximation of $\mathrm{E}(\widetilde{S}|\overline{q})$ with 100 samples), we established that it is the deterministic part that is responsible for the improvement (not shown).} \sout{We leave the investigation of this physical mechanism for future research.} 
Note that generative models (GAN and VAE) accurately reproduce kinetic energy growth in transitional flow (panel (a), $t < 2$ years) in agreement with the offline analysis. 

\begin{figure} [h!]
\centering
\includegraphics[width=1\textwidth]{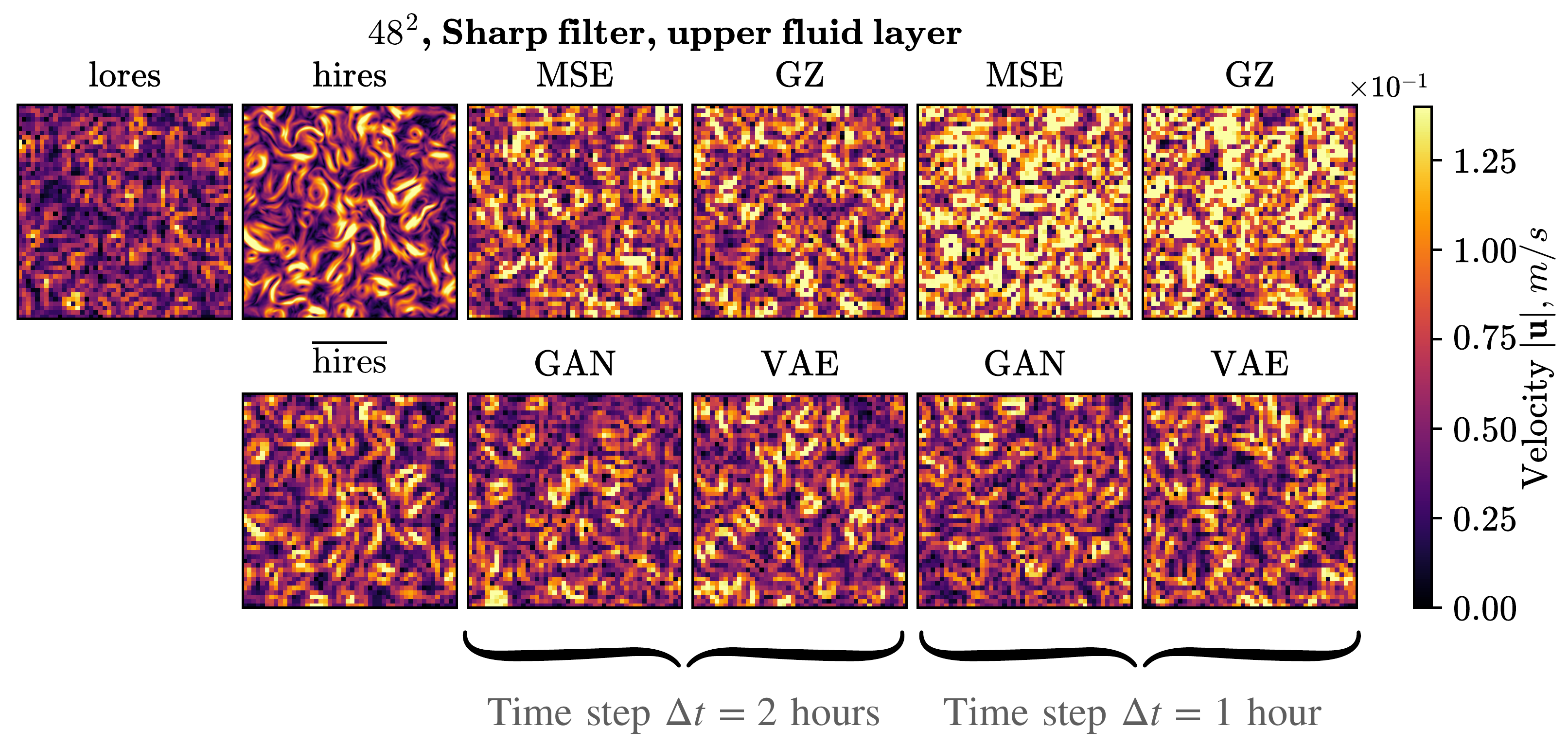}
\caption{Snapshots of the modulus of velocity. Two columns with $\Delta t = 2$ hours correspond to Figure \ref{fig:online_48}. The smaller the time step, the smaller the effective eddy viscosity, see \ref{appendix:online}.}
\label{fig:snapshot}
\end{figure}

\begin{figure} [h!]
\centering
\includegraphics[width=1\textwidth]{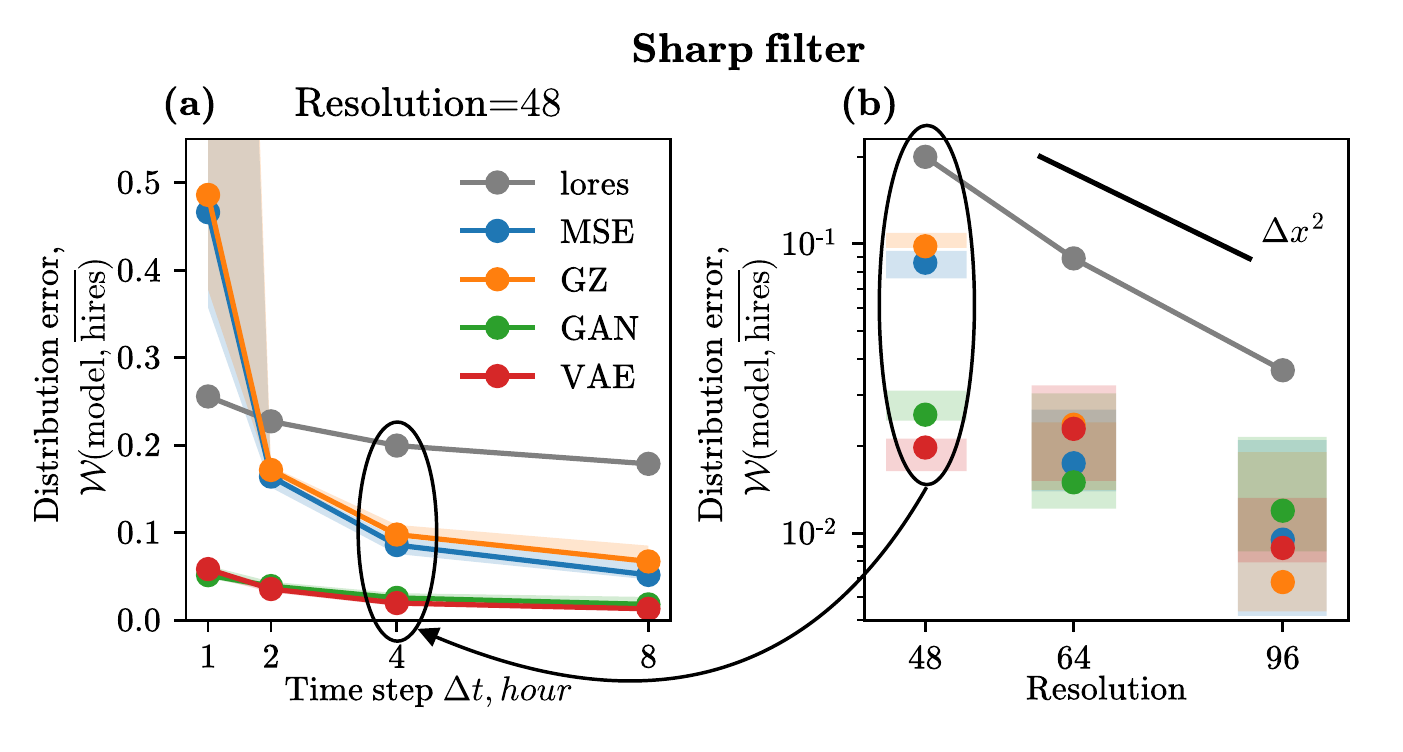}
\caption{Online distributional metric (Eq. \eqref{eq:distribution_metric}): (a) as a function of time step at the coarsest spatial resolution and (b) as a function of spatial resolution. The shading area shows min-max values
among training realizations, and markers show median value. The smaller the time step, the smaller the effective eddy viscosity, see \ref{appendix:online}.}
\label{fig:distribution_error}
\end{figure}

Snapshots of the velocity modulus are shown in Figure \ref{fig:snapshot}. At time step $\Delta t = 2$ hours baseline models (MSE and GZ) have too many small eddies, and at time step $\Delta t = 1$ hour the flow becomes unphysical and overenergized. GAN and VAE models at both time steps produce physical solutions which look similar to the filtered and coarse-grained high-resolution simulation ($\overline{\mathrm{hires}}$). In Figure \ref{fig:distribution_error}(a) we show distributional metric as a function of the time step. While baseline models (MSE and GZ) are very sensitive to the time step, the generative models (GAN and VAE) are relatively insensitive to the time step and have the smallest errors. This suggests that the generative stochastic models have better numerical stability properties. See \ref{appendix:online} for further discussion on numerical stability. 
%In \ref{appendix:online} we explain that a small time step corresponds to the smaller effective eddy viscosity, and our statement about numerical stability becomes more evident. 

In Figure \ref{fig:distribution_error}(b) we show the distributional metric as a function of resolution. At the coarsest resolution $48^2$, the generative stochastic models (GAN and VAE) have 5--10 times lower error compared to the unparameterized simulation (lores) and 3--5 times lower error compared to the baseline models (GZ and MSE). 
%\sout{The error for generative models at $48^2$ is comparable to the error of unparameterized simulation with twice the resolution $96^2$. We emphasize that it is an important result: LES models should be able to predict instant or statistical quantities as good as higher resolution models to make LES approach practically applicable \cite{kochkov2021machine}. However, this result should be understood with caution: an unparameterized model at resolution $96^2$ resolves much more turbulent eddies and filaments, which cannot be resolved on a coarse grid.} 
For intermediate and higher resolutions ($64^2$ and $96^2$) all ML-based models (GZ, MSE, GAN, VAE) improve distributional error compared to the unparameterized model, but the confidence intervals (shading area) exceed the difference between the median values. So we conclude that the effect of stochastic subgrid models (GZ, GAN, VAE), as opposed to the deterministic one (MSE), at these resolutions is negligible. The discrepancy between the offline and online analysis may be due to the inclusion of $ssd$ term, time sampling method of the stochastic parameterization, and time integration scheme. Overall, generative models (GAN and VAE) improve simulation if there are issues with numerical stability, and perform as well as the baseline deterministic model in other cases. \textcolor{red}%{Note that many of the presented data-driven parameterizations outperform the unparameterized higher-resolution model according to the online metric.} 
The spectral-error metric reported in \ref{appendix:online} yields similar conclusions. %with respect to an additional online metric (spectral error, \ref{appendix:online}) are the same.

\begin{figure} [h!]
\centering
\includegraphics[width=1\textwidth]{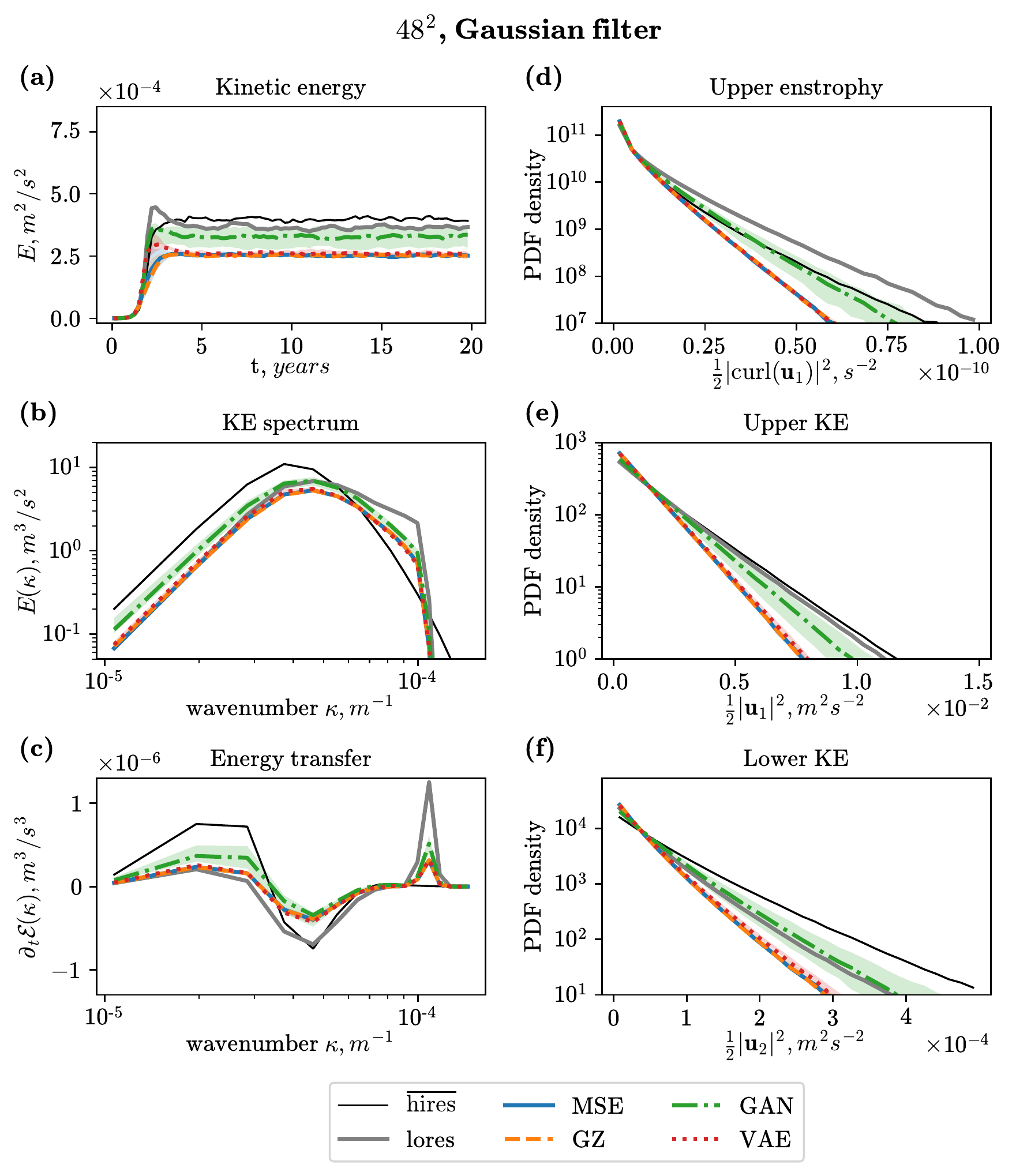}
\caption{Online simulations with parameterized models. Similar to Figure \ref{fig:online_48}, but for models trained on the dataset with Gaussian filter.}
\label{fig:online_48_gauss}
\end{figure}

In Figure \ref{fig:online_48_gauss} we show the online results for the subgrid models trained on the dataset produced using the Gaussian filter. The subgrid models cannot substantially improve the KE spectrum on large scales with respect to the unparameterized model (panel (b)), and it results in little or no improvement in the other statistical characteristics. 
At higher resolutions ($64^2$ and $96^2$) we observe the improvement in reproducing the KE spectrum on small scales, but not the large ones (not shown). Similar to \citeA{zanna2020data}, we report in \ref{appendix:online} how the kinetic energy in online simulation changes when the subgrid model is multiplied by the adjustable parameter. This characteristic clearly demonstrates that subgrid models trained for the Gaussian filter are less efficient in energizing the flow. The same issues for the models trained to predict the subgrid forcing diagnosed with the Gaussian filter were reported in \citeA{ross2022benchmarking} and these may be caused by the mentioned discrepancies between the offline and online analysis.

In \ref{appendix:online} we include additional online results. The online generalization to the turbulence configuration with jets shows that generative models clearly improve the simulation of the transitional flow, but at a later time, all the subgrid models including baselines experience numerical stability issues. Runtime for the generative models is the same as for the deterministic baseline.

\clearpage

\section{Conclusions and discussion}

In this work, we propose to leverage generative machine-learning models (GAN and VAE) to build stochastic subgrid parameterizations of mesoscale eddies. Generative models allow to sample from the conditional distribution of subgrid forcing given resolved variables. We performed offline and online evaluations of the proposed subgrid models, and compared them against baseline deterministic and stochastic ML models in an idealized ocean simulation for a range of resolutions. 
%We set the architecture of generative models such as to perform a fair comparison to a recently proposed stochastic ML subgrid model (GZ, \citeA{guillaumin2021stochastic}), which we consider as a baseline. We train all stochastic models on the same dataset and with the same building block used for inference -- the convolutional neural network.

Our main findings can be summarized as follows:
\begin{itemize}
    %\item All stochastic models produce a flow-dependent estimation of the uncertainty as it is seen by the spatial structure of the field of conditional standard deviation.
    %\item The stochastic residuals of generative models are spatially correlated. We did not specify it as the target property during training, and thus we expect that generative models could represent other statistically significant properties of subgrid forcing in more realistic settings.
    %\item The stochastic backscatter is concentrated near the grid cut-off and generative models attempt to better simulate this process by introducing spatial correlation.
    \item Generative models are able to simulate the stochastic residuals of subgrid forcing with spatial structure similar to the true residuals.
    \item Generative models accurately represent the energy transfer spectrum and thus reproduce the large-scale kinetic energy backscatter missing at coarse-resolution.
    % \item GAN model improves the power spectrum of subgrid forcing compared to the GZ model and its deterministic part.
    \item The GAN model is superior to others according to the offline metrics for subgrid forcing; however, the VAE model demonstrates better offline generalization to the unseen turbulence configuration (meandering jets).
    %\item Generative models are more accurate in transitional turbulence regime ($t<2$ years) both offline and online.
    \item Both generative models (GAN and VAE) improve the numerical stability properties and prevent overamplification of the unphysical flows in online simulations at the coarsest resolution compared to the baseline ML models. 
    %This effect is present for the coarsest resolution. For finer resolutions, generative models perform as well as the baseline.
    %\item \textcolor{red}{Many of the presented stochastic parameterizations outperform an unparameterized higher-resolution model in online simulations according to the distributional and spectral metrics.}
\end{itemize}

%We observed two major difficulties during the training of generative models: suboptimal prediction of the deterministic part (conditional mean) and insufficient spread of the conditional distribution. 
%The problem of insufficient spread for GAN model was resolved by choosing the loss function suggested in \citeA{adler2018deep}, but for the VAE model, the spread remains underestimated after all our trials to change the loss function or architecture of the neural network. In spite of the small spread, VAE model is able to generate stochastic residuals with proper spatial structure. Both generative models (GAN and VAE) deteriorate the conditional mean prediction evaluated by the RMSE metric compared to the baseline model GZ. However, we note that RMSE is still better than we expect for physically-based parameterizations.%, and generative models are still able to simulate the average energy transfer, i.e. large-scale kinetic energy backscatter. The less important difficulties with training were: an absence of clear stopping criteria for training generative models and models with uncommon behavior, see \ref{appendix:offline}.

In spite of the different performance of GAN and VAE models in the offline analysis, their performance is  similar in online simulations. 
Therefore, offline metrics or loss functions may be bad proxies for the online performance \cite{frezat2022posteriori,ross2022benchmarking}. 
The energy transfer spectrum is one of the main properties of subgrid forcing which is essential to properly energize the flow and could be considered as an alternative loss function. However, the spatial structure of the subgrid forcing and stochastic residuals may be important to ensure the development of the physical solution.

%\textcolor{red}{The generative approach can suggest alternative loss functions which may result to subgrid models with better properties.} 

%The only property that we feel is of absolute importance is the spectrum of average energy transfer: subgrid forcing diagnosed with the Gaussian filter has a "wrong"{} energy transfer and we do not observe substantial improvement by any of the proposed subgrid models. 

Our online simulations are optimal when the time correlation of the latent variable sampling is equal to the model timestep, which is equivalent to a white noise model and consistent with our offline training methodology. 
The effect of the parameterization can be analyzed by decomposing it into deterministic and stochastic parts. 
The determinitic part is defined as the conditional mean; while the stochastic part as a white noise model. 
The white noise process model implies that the energy injection by the stochastic part of the parameterization approaches zero in the limit of the small time steps \cite{alvelius1999random}. 
In addition, the average energy injection is fully described by the deterministic part of the parameterization (i.e., conditional mean, see \citeA{moser2021statistical}). %We believe that 
%it is possible to reveal the role of the time sampling method and 
One can modify the definition of the subgrid model, for example by including memory effects, to generate a stochastic model with non-vanishing energy input \cite{chorin2015discrete, gagne2020machine, agarwal2021comparison,delsole2000fundamental, berner2005linking, bhouri2022history}.

The important property of the proposed generative models:  they do not introduce new limitations to be trained on the global ocean data compared to our baseline model (GZ). Moreover, compared to GZ, both generative models do not require an explicit expression for the likelihood function, and thus slightly more complicated architecture of the stochastic model can be used, for example, the final divergence layer \cite{zanna2020data} which allows building conservative parameterizations. We expect that the application of generative models for complex flows may greatly improve the quality of the generated stochastic residuals compared to the traditional methods.

\clearpage
\appendix
\section{Numerical schemes and small-scale dissipation} \label{appendix:numerical_schemes}
Integration in time is performed with third-order Adams-Bashford scheme \cite{durran1991third}.
Equations \eqref{eq:gov_eq_1}, \eqref{eq:gov_eq_2} are approximated in space on a uniform collocated grid with the use of the pseudo-spectral method, i.e. all derivatives are computed in Fourier space, but the only nonlinear operation $(\mathbf{u}_m q_m)$ is computed in physical space \cite{fox1973pseudospectral}. 
%Usually, this spatial approximation is used together with 2/3--dealiasing procedure \cite{orszag1971elimination}, which retains 2/3 of the lowest spatial frequencies and filters out 1/3 of the highest spatial frequencies after each time step. It allows to fulfill quadratic conservation laws of energy and potential enstrophy inherent to continuous advection operator. 

In our numerical solver, aliasing errors are reduced with the use of "exponential cut-off"{} filter \cite{canuto2012spectral} denoted as $ssd$, which attenuates highest spatial frequencies and additionally removes enstrophy near the grid scale thus ensuring numerical stability. Application of the filter $ssd(\cdot)$ is equivalent to multiplication by the following function in Fourier space:
\begin{equation}
        \widehat{ssd}(k,l) = 
            \begin{cases}
                1, &\kappa < \kappa_c \\
                \exp (-23.6 (\Delta x)^4 (\kappa - \kappa_c)^4), & \text{otherwise}
            \end{cases} \label{eq:ssd_filter}
    \end{equation}
where $\kappa = \sqrt{k^2+l^2}$ is radial wavenumber,
$k$ and $l$ are zonal and meridional wavenumbers, respectively,
    $\Delta x$ is grid step of the model, 
    $\kappa_{max}=\frac{\pi}{\Delta x}$ is maximum
    wavenumber, $\kappa_c = 0.65 \kappa_{max}$.
We simplify notation when placing $ssd$ into the right-hand side of the  equation \eqref{eq:gov_eq_1}. Instead, every time step from layer $n$ to a new layer ($q_m^n \rightarrow q_m^*$) is followed by the application of the filter $q_m^{n+1}=ssd(q_m^*)$ \cite{lacasce1996baroclinic, arbic2003coherent}. 

We do not add molecular viscosity into governing equations \eqref{eq:gov_eq_1}, and thus formally have "infinite"{} Reynolds number. However, the dissipation is provided by the term $ssd$ which depends on the grid step (see Eq. \eqref{eq:ssd_filter}). Grid-dependent small-scale dissipation is a typical choice in ocean modeling \cite{griffies2000biharmonic}. Dissipation introduced by $ssd$ is relatively small, which is important for the simulation of quasi-two dimensional turbulence \cite{thuburn2014cascades}. An example of the undesirable effect of the dissipative model is shown in Figure \ref{fig:fig1}(d), where $ssd$ is balanced with energy transfer, and thus induces spurious forward energy cascade.

\section{LES filters} \label{appendix:filters}
Spatial filter  $\overline{(\cdot)}$  (Eq. \eqref{eq:spatial_filter}) consists of two operations: spectral coarsegraining which reduces the resolution of the image and spectral smoothing. Because both are defined as a pointwise function in Fourier space, they commute and can be composed into a single operator, which we often call a "filter"{}.

Filter $\overline{(\cdot)}$ is applied as multiplication in Fourier space by the following function:
\begin{itemize}
    \item "Gaussian"{}
    \begin{equation}
        \widehat{G}(k,l) = 
        \begin{cases}
            \exp (-\kappa^2 (2 \Delta x)^2 / 24), & \text{otherwise} \\
            0, & \kappa_{max} \leq k \text{ or } \kappa_{max} \leq l
        \end{cases} \label{eq:gauss_filter}
    \end{equation}
    Where $\kappa_{max}=\frac{\pi}{\Delta x}$ and $\Delta x$ is the grid step of the coarse model. After discarding the frequencies above $\kappa_{max}$, the filtered signal is represented on a coarse mesh \cite{ghosal1996analysis}.
    According to the definition of filter width given by \citeA{lund1997use}, the width of this Gaussian
    filter is $2 \Delta x$, which is twice as large as the grid step of the coarse model.
    
    \item "Sharp"{}
    \begin{equation}
        \widehat{G}(k,l) = 
            \begin{cases}
                1, &\kappa < \kappa_c \\
                \exp (-23.6 (\Delta x)^4 (\kappa - \kappa_c)^4), & \text{otherwise} \\
                0, & \kappa_{max} \leq k \text{ or } \kappa_{max} \leq l
            \end{cases} \label{eq:model_filter}
    \end{equation}
    where $\kappa_c = 0.65 \kappa_{max}$. This filter is given
    by a combination of sharp cut-off coarsegraining
    and model filter (Eq. \eqref{eq:ssd_filter}).
\end{itemize}
Motivation for using these filters is given in \citeA{ross2022benchmarking}.

\section{Training of the Machine Learning Models}
\label{appendix:training_information}
\subsection{GAN loss function}
The presented below training algorithm closely resembles paper of \citeA{adler2018deep}, where the discriminator analyzes two generated images.

We generate two images $\widetilde{S}_1=G(z_1, \overline{q})$ and $\widetilde{S}_2=G(z_2, \overline{q})$ for a given $\overline{q}$ with two samples from standard normal distribution $z_1, z_2 \in \mathbb{R}^{2\times n \times n}$ and stack them in layer dimension: %introduce new variables stacked in layer dimension:
 \begin{equation*}
     \mathbf{S}_1 = (\widetilde{S}_1, S), ~
     \mathbf{S}_2 = (S, \widetilde{S}_2), ~
     \widetilde{\mathbf{S}} = (\widetilde{S}_1, \widetilde{S}_2),
 \end{equation*}
 where $\mathbf{S}_1, \mathbf{S}_2, \widetilde{\mathbf{S}} \in \mathbb{R}^{4 \times n \times n}$. The WGAN loss (Eq.
 \eqref{eq:GAN_loss_simple}) for a single data sample transforms to:
\begin{equation*}
    %\max_{D \in \mathrm{Lip}(\mathbf{S})} \mathbb{E} 
    %\mathcal{L}_{\mathrm{W}}, \text{ where}\\
    \mathcal{L}_{\mathrm{W}} = \left[ \frac{1}{2} \left( D(\mathbf{S}_1, \overline{q}) + D(\mathbf{S}_2, \overline{q}) \right) - D(\widetilde{\mathbf{S}}, \overline{q}) \right].
\end{equation*}
 The discriminator $D$ should be 1-Lipschitz in the first argument, and we enforce it with the gradient penalty (WGAN-GP, \citeA{gulrajani2017improved}):
 %is 1-Lipschitz on stacked variable denoted as $\mathrm{Lip}(\mathbf{S})$. Gradient penalty \cite{gulrajani2017improved} is a regularization which removes constraint $D \in \mathrm{Lip}(\mathbf{S})$:
 \begin{equation*}
     \mathcal{L}_{\mathrm{grad}} = \left( ||\nabla_{\widehat{\mathbf{S}}} D(\widehat{\mathbf{S}},\overline{q})||_2-1 \right)^2,
 \end{equation*}
 where $\widehat{\mathbf{S}}=\epsilon \mathbf{S} + (1-\epsilon)\widetilde{\mathbf{S}}$. The random number $\epsilon$ is uniformly distributed on $[0,1]$ and chosen uniquely for every training sample. For every batch we choose randomly $\mathbf{S}$ from set $\{\mathbf{S}_1, \mathbf{S}_2\}$. Note that $D(\widehat{\mathbf{S}},\overline{q}) \in \mathbb{R}$, $\nabla_{\widehat{\mathbf{S}}} D(\widehat{\mathbf{S}},\overline{q}) \in \mathbb{R}^{4 \times n \times n}$ and norm $||\cdot||_2$ for tensor is defined above. Regularization preventing drift of discriminator:
 \begin{equation*}
     \mathcal{L}_{\mathrm{drift}} = \left[ D(\mathbf{S}_1, \overline{q}) \right]^2.
 \end{equation*}

 We minimize the following loss for the discriminator:
 \begin{equation}
     \mathcal{L}_{\mathrm{D}} = - \mathcal{L}_{\mathrm{W}} + 10 \mathcal{L}_{\mathrm{grad}} + 10^{-3} \mathcal{L}_{\mathrm{drift}}, \label{eq:D_loss}
 \end{equation}
and the loss to be minimized for the generator is:
\begin{equation}
    \mathcal{L}_{\mathrm{G}} = - D(\widetilde{\mathbf{S}}, \overline{q}). \label{eq:G_loss}
\end{equation}
In the original paper  $\mathcal{L}_{\mathrm{G}}=\mathcal{L}_{\mathrm{W}}$ \cite{adler2018deep}, but we follow a typical approach when only generated samples constitute the generator loss \cite{dong2019towards}.

Discriminator $D$ is parameterized by DCGAN discriminator \cite{radford2015unsupervised} with two modifications: we remove the activation function in the final layer and remove batch normalization because it is necessary for proper use of gradient penalty \cite{gulrajani2017improved}.
Following \citeA{arjovsky2017wasserstein}, we optimize the discriminator loss (Eq. \eqref{eq:D_loss}) for five batches in a row, and then we optimize the generator loss (Eq. \eqref{eq:G_loss}) for one batch.

\subsection{VAE loss function}
To train the VAE model we parameterize every probability density in the VAE loss function (Eq. \eqref{eq:vae_loss}) with Gaussian distributions.

The distributions for encoder, decoder and prior, respectively:
\begin{gather}
    q_{\phi}(z|S,\overline{q})=\mathcal{N} \left(\mu_{\phi}(S,\overline{q}), \mathrm{diag}(\sigma^2_{\phi}(S,\overline{q})) \right) \\
    \rho_{\theta}(S|z, \overline{q}) = \mathcal{N} \left(\mu_{\theta}(z,\overline{q}), \gamma I \right) \\
    \rho(z)=\mathcal{N}(0,I),
\end{gather}
where $I \in \mathcal{R}^{2n^2 \times 2n^2}$ is identity matrix, $\gamma$ is free parameter and $S,\overline{q},z \in \mathbb{R}^{2 \times n \times n}$. The mappings $\mu_{\phi}$, $\ln(\sigma^2_{\phi})$ and $\mu_{\theta}$ are deterministic. 

The loss function to be minimized (Eq. \eqref{eq:vae_loss}) for one training sample transforms to:
\begin{equation}
    \mathcal{L}_{\mathrm{VAE}} = \frac{1}{2 \gamma} ||S - \mu_{\theta}(\hat{z},\overline{q}) ||^2_2 + \frac{1}{2}\sum_{m,i,j} \left( \sigma_{\phi}^2 + \mu_{\phi}^2 -1 -\ln ( \sigma^2_{\phi})  \right)_{m,i,j},
\end{equation}
where $\hat{z}$ is one sample from encoder distribution, i.e. $\hat{z}=\mu_{\phi} + \epsilon \sigma_{\phi}$, $\epsilon\sim\mathcal{N}(0,I)$. Note that as suggested by \citeA{rybkin2021simple}, we sum values of MSE loss and KL loss across dimensions. The variance of decoder distribution $\gamma$ is a parameter regulating the relative importance of reconstruction and regularization terms. According to \citeA{takida2022preventing}, common problems of VAE such as posterior collapse and smoothness of generated images may result from the incorrect choice of parameter $\gamma$. Following \citeA{rybkin2021simple}, we estimate the variance of the decoder as a mean squared error: $\gamma=\frac{1}{2 n^2}||S - \mu_{\theta}||^2_2$. We compute $\gamma$ uniquely for every batch and do not differentiate it. %See \ref{appendix:training_information} for additional training information.

\subsection{Additional training information} 
\label{appendix:specific_training_information}
All image-to-image mappings (mean and variance prediction in GZ, generator in GAN, encoder and decoder in VAE) are based on the same convolutional neural network (CNN) similar to \citeA{guillaumin2021stochastic, ross2022benchmarking} with parameters given in Table \ref{tab:CNN}. We follow a common approach with the normalization of input and output variables before passing them to neural networks. Each channel representing a different physical quantity or different fluid layer is normalized by a unique standard deviation computed over the training dataset. Note that the variance channel of GZ model is normalized by the squared standard deviation of the mean channel. Normalization constants become part of the model and they are not adjusted in offline or online tests.
%We do not distinguish explicitly between normalized and unnormalized quantities in the main text, but follow a common convention: only loss functions operate on normalized data, while metrics operate on unnormalized data.

Models are trained in Pytorch \cite{paszke2019pytorch}, batch size is $64$, training algorithm is Adam \cite{kingma2014adam} with standard parameters $(\beta_1, \beta_2)=(0.9,0.999)$ for GZ and VAE, and $(\beta_1, \beta_2)=(0.5,0.999)$ for GAN \cite{radford2015unsupervised}. The learning rate is $lr=0.001$ for GZ and $lr=0.0002$ for GAN and VAE. GAN and VAE models are optimized for 200 epochs, and in GZ model each channel (mean and variance) is optimized for 50 epochs. Early stopping or any other criteria for choosing the best epoch was not used. Weight decay was not used. We use the following scheduler of the learning rate for GZ and VAE: on every milestone $[1/2,3/4,7/8]\cdot N_{\mathrm{epoch}}$ multiply learning rate by $\gamma=0.1$, for GAN $\gamma=0.5$. Weights of the discriminator and generator of GAN are initialized with zero mean and standard deviation $0.02$ \cite{radford2015unsupervised}. During inference, neural networks are switched to evaluation mode so that batch normalization layers use parameters accumulated during training. 

\begin{table}[h!]
 \caption{Configuration of convolutional neural network (CNN) parameterizing image to image mapping.}
 \centering
 \begin{tabular}{l| l}
    Number of input/output images & arbitrary ($n_{in}$, $n_{out}$)\\
    Resolution of input/output/hidden layers & arbitrary, but the same \\
    Number of filters & $128$, $64$, $32$, $32$, $32$, $32$, $32$, $n_{out}$ \\
    Filter width & $5$, $5$, $3$, $3$, $3$, $3$, $3$, $3$ \\
    Boundary conditions & periodic ("circular padding"{}) \\
    Activation function & ReLU, in hidden layers \\
    Batch normalization & after ReLU, in hidden layers
\end{tabular}
\label{tab:CNN}
\end{table}

\clearpage

\section{Additional offline results} \label{appendix:offline}
\begin{figure}[h!]
\includegraphics[width=\textwidth]{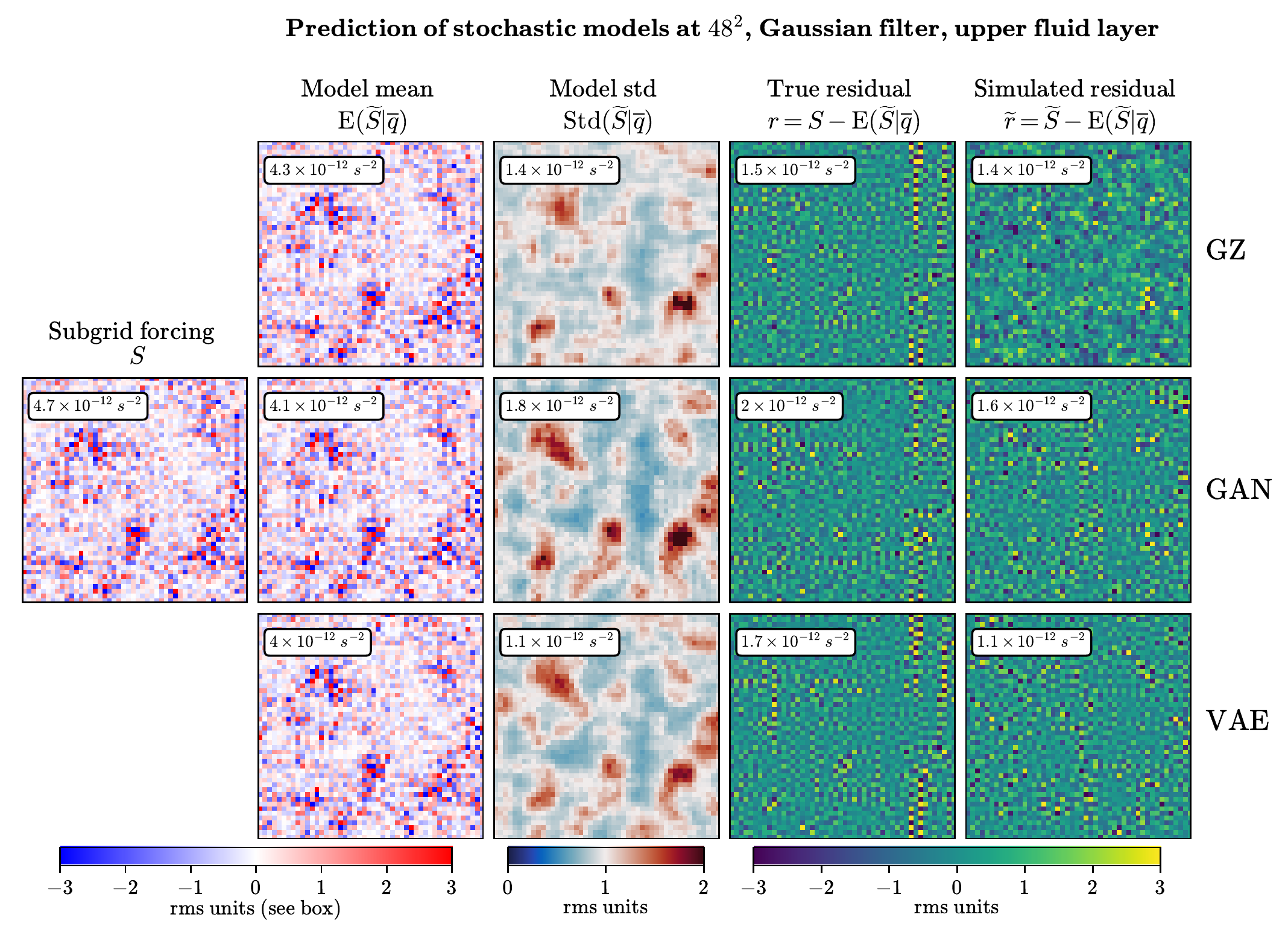}
\caption{Prediction of subgrid forcing on the testing dataset. Same as Figure \ref{fig:SGS_snapshot} but for Gaussian filter.}
\label{fig:D1}
\end{figure}

\begin{figure}[h!]
\centering
\includegraphics[width=1\textwidth]{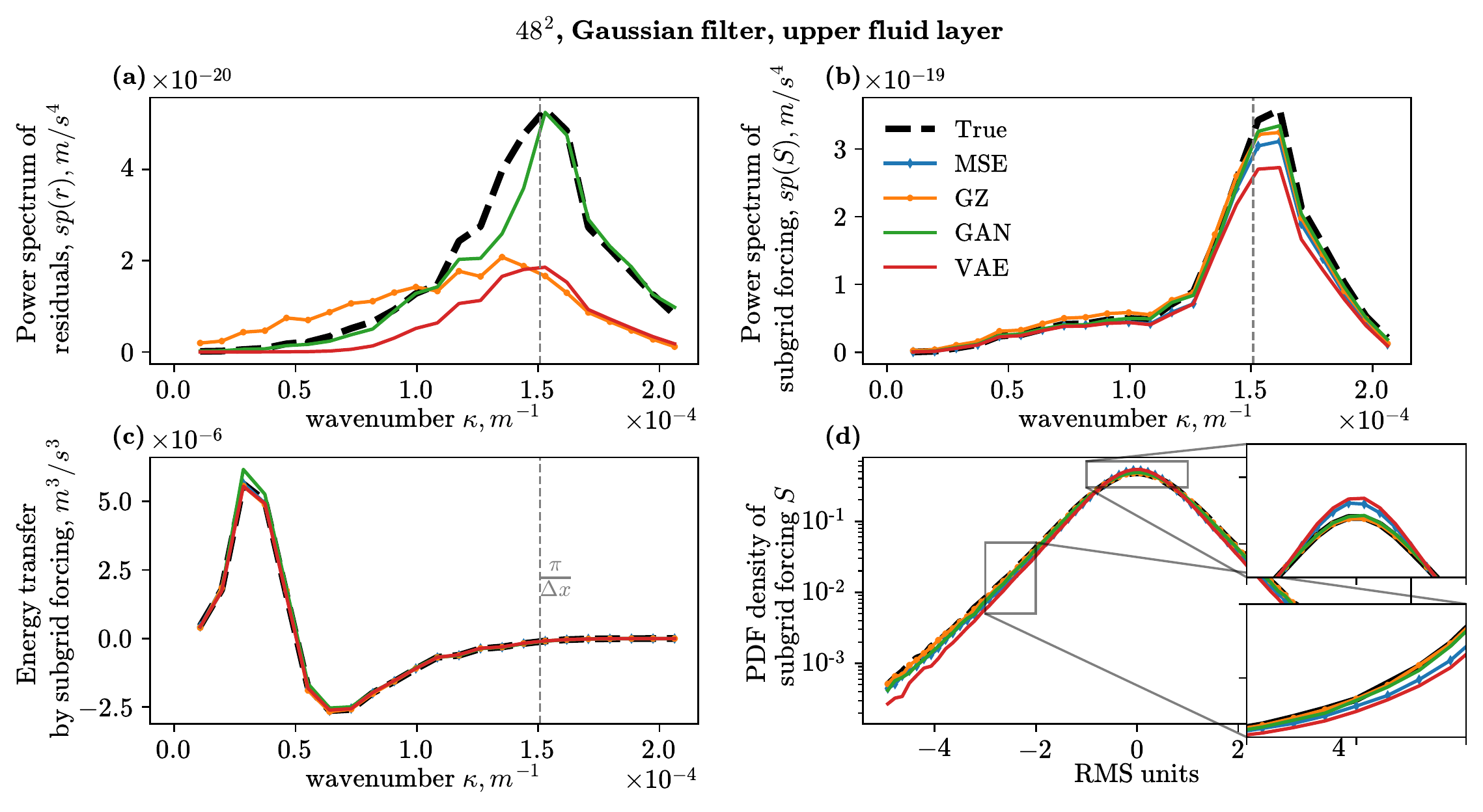}
\caption{Offline analysis of subgrid models. Same as Figure \ref{fig:spectra_offline} but for Gaussian filter.}
\label{fig:D2}
\end{figure}

\begin{figure} [h!]
\centering
\includegraphics[width=1\textwidth]{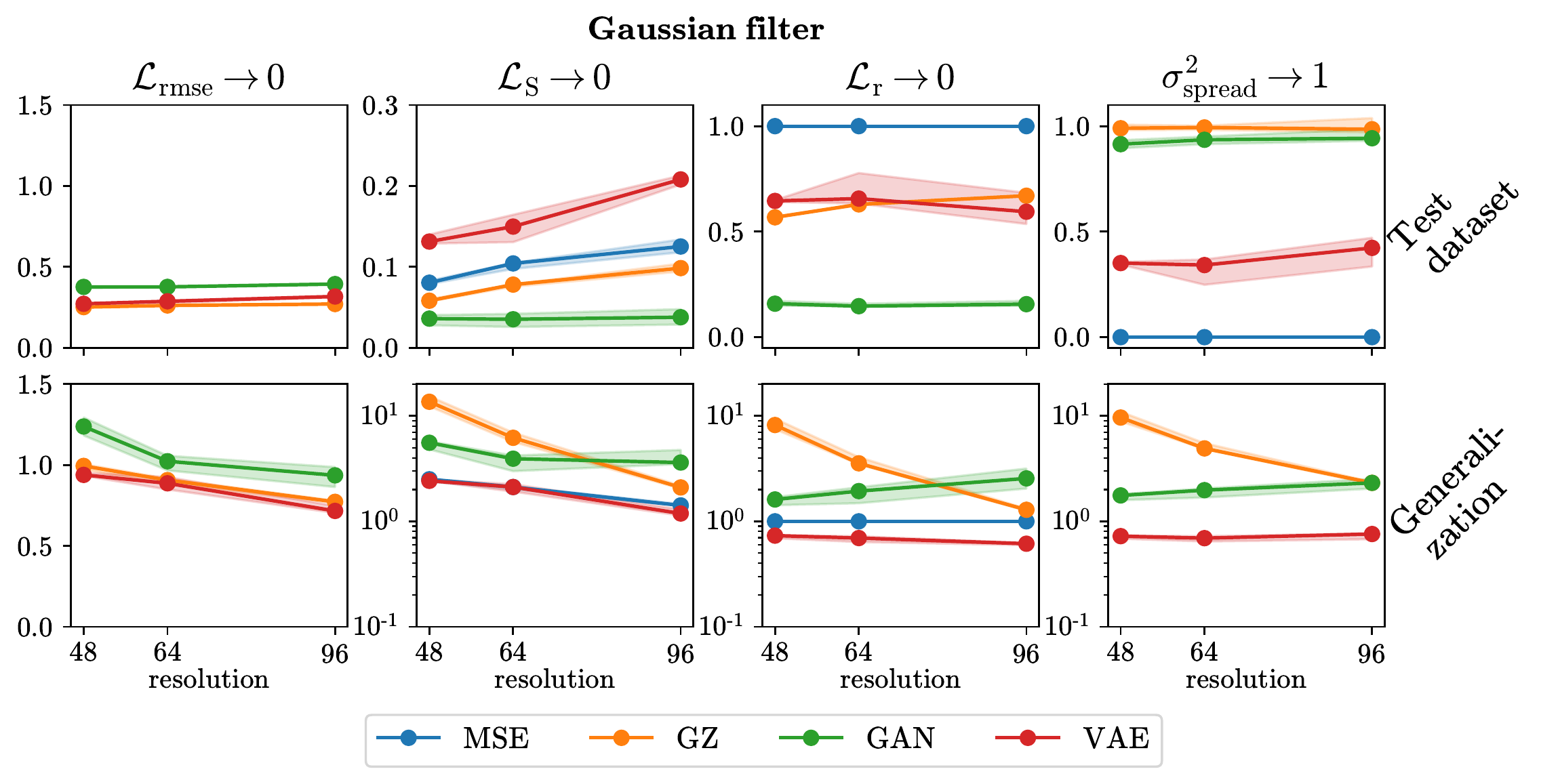}
\caption{Offline metrics from Table \ref{tab:offline_metrics}. Same as Figure \ref{fig:offline_metrics} but for Gaussian filter.}
\label{fig:D3}
\end{figure}

In Figures \ref{fig:D1}, \ref{fig:D2} and \ref{fig:D3} we show the results of offline analysis for the Gaussian filter. Our main conclusions about the performance of the stochastic models are the same as for the Sharp filter.

% In Figure \ref{fig:offline_metrics_sharp} we show offline metrics, but for models trained and evaluated on the dataset with Sharp filter. General conclusions about the offline performance of stochastic models are the same as for the Gaussian filter. The major difference is in the three models, which demonstrate untypical behavior. In particular, the VAE model at resolution $64^2$ in two training realizations out of 5 experiences a posterior collapse problem, i.e. spread is zero $\sigma^2_{\mathrm{spread}} \approx 0$. This problem is not serious: it can be detected during training by evaluating the spread metric on the validation dataset and can be resolved by choosing a different random seed. A more serious problem was observed with the GZ model in one training realization out of 5 at resolution $48^2$. This GZ model has typical metrics on the testing dataset, but errors ($\mathcal{L}_{\mathrm{rmse}}$ and $\mathcal{L}_{\mathrm{S}}$) on generalization dataset are very large, see the orange diamond. Note that it is not a result of the usage of a complicated loss function, but it happened to the deterministic channel of the GZ model trained with MSE loss. Generalization issues can be overcome by choosing special architecture of neural networks \cite{frezat2021physical, guan2022learning, pawar2022frame}, input-output variables \cite{ross2022benchmarking} or including diverse dynamical regimes into dataset \cite{guillaumin2021stochastic, beucler2021climate}, but it is not the objective of our study. We simply exclude outliers from the subsequent online analysis.

\clearpage

\section{Additional online results} 
\label{appendix:online}
\begin{figure} [h!]
\centering
\includegraphics[width=0.6\textwidth]{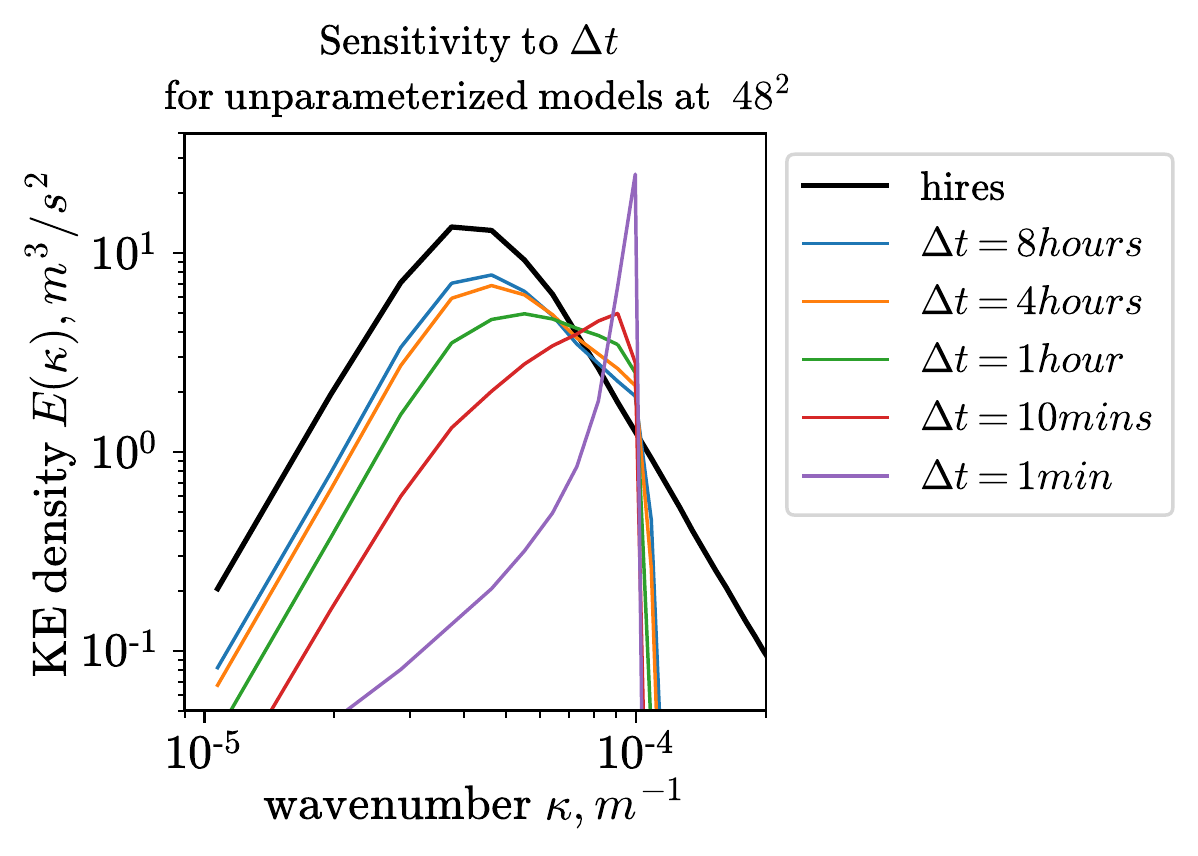}
\caption{Here we show that the smaller the time step, the smaller the effective eddy viscosity and the closer the unparameterized coarse model to inviscid simulation, i.e. energy accumulates near the cut-off for a small time step.}
\label{fig:sensitivity_to_time_step}
\end{figure}
Below we show that the time step is connected to the effective eddy viscosity in our particular numerical scheme, and thus the sensitivity of the online simulation results to the time step reveals the numerical stability properties. Small-scale dissipation ($ssd$) was formulated not as a tendency in RHS of the governing equation, but as a postprocessing operation following every time step, see \ref{appendix:numerical_schemes}. Because  $ssd$ does not contain a time step explicitly, the effect of dissipation accumulates over several time steps \cite{lund2003use}. An effective filter that we apply to the solution per unit time interval (if solution is steady) is $\widehat{ssd}^{1/\Delta t}$, which converges at every radial wavenumber $\kappa$ to Heaviside step function 
\begin{equation}
    (\widehat{ssd}(\kappa))^{1/\Delta t} \to H(0.65 \kappa_{max} - \kappa) \text{ as } \Delta t \to 0.
\end{equation}
 Filtering with this Heaviside step function approximately corresponds to 2/3--dealiasing scheme \cite{orszag1971elimination}, which is the discretization of inviscid equations and conserves energy and enstrophy. So, we expect that the smaller the time step, the smaller the effective eddy viscosity produced by the $ssd$ term. By refining the time step, we established that unparameterized coarse models behave as inviscid simulation at $\Delta t\to 0$, i.e. energy accumulates near the grid scale, see Figure \ref{fig:sensitivity_to_time_step}.

Below we derive a spectral metric for the analysis of online simulations. The rate of change of total energy (Eq. \eqref{eq:E_definition}) is defined as
\begin{equation}
    \partial_t \mathcal{E} = -\frac{1}{H} \sum_{m=1}^{2} H_m  \langle \psi_m \partial_t q_m \rangle.
\end{equation}
Applying this formula to the governing equation \eqref{eq:gov_eq_1} and using Parseval theorem, the rate of change of total energy in Fourier space is
\begin{equation}
    \begin{split}
    \partial_t \mathcal{E}(k,l) = 
    - \frac{1}{H} \sum_{m=1}^{2} H_m  Re \Bigg( & \widehat{\psi}_m^* \partial_t \widehat{q}_m \Bigg)=  \\
    \frac{1}{H}\sum_{m=1}^{2} H_m Re\Bigg( &
    \underbrace{\widehat{\psi}_m^* \widehat{\nabla(\mathbf{u}_m q_m})}_{\text{energy transfer}} +
    \underbrace{\widehat{\psi}_m^* U_m \widehat{\partial_x q}_m}_{\text{energy source}} + 
    \underbrace{\widehat{\psi}_m^* \delta_{m,2} r_{ek} \widehat{ \nabla^2 \psi}_m}_{\text{energy dissipation}}
    \Bigg).
    \end{split} \label{eq:energy_balance_unfiltered}
\end{equation}
Here we neglected the contribution from the small-scale dissipation term $ssd$ which acts on a limited set of wavenumbers. The corresponding energy balance equation for filtered and coarsegrained system \eqref{eq:les_eq_1} is
\begin{equation}
    \begin{split}
    \partial_t \overline{\mathcal{E}}(k,l) = 
    - \frac{1}{H} \sum_{m=1}^{2} H_m  Re \Bigg( & \widehat{\overline{\psi}}_m^* \partial_t \widehat{\overline{q}}_m \Bigg)=  \\
    \frac{1}{H}\sum_{m=1}^{2} H_m Re\Bigg( &
    \underbrace{\widehat{\overline{\psi}}_m^* \widehat{\nabla(\overline{\mathbf{u}_m q_m})}}_{\text{energy transfer}} +
    \underbrace{\widehat{\overline{\psi}}_m^* U_m \widehat{\partial_x \overline{q}}_m}_{\text{energy source}} + 
    \underbrace{\widehat{\overline{\psi}}_m^* \delta_{m,2} r_{ek} \widehat{ \nabla^2 \overline{\psi}}_m}_{\text{energy dissipation}}
    \Bigg).
    \end{split} \label{eq:energy_balance}
\end{equation}
 In Eq. \eqref{eq:energy_balance} the energy transfer is proportional to $\sim \overline{\mathbf{u}_m q_m}$ and can be split into the resolved transfer $\sim \overline{\mathbf{u}}_m \overline{q}_m$ and unresolved transfer $\sim \overline{\mathbf{u}_m q_m} - \overline{\mathbf{u}}_m \overline{q}_m$ which is parameterized by the subgrid model. Every term in equation \eqref{eq:energy_balance} can be obtained from the corresponding term in the energy balance of high-resolution simulation (Eq. \eqref{eq:energy_balance_unfiltered}) by multiplying twice by the filter transfer function, i.e. by $(\widehat{G}(k,l))^2$.

\begin{figure} [h!]
\centering
\includegraphics[width=1\textwidth]{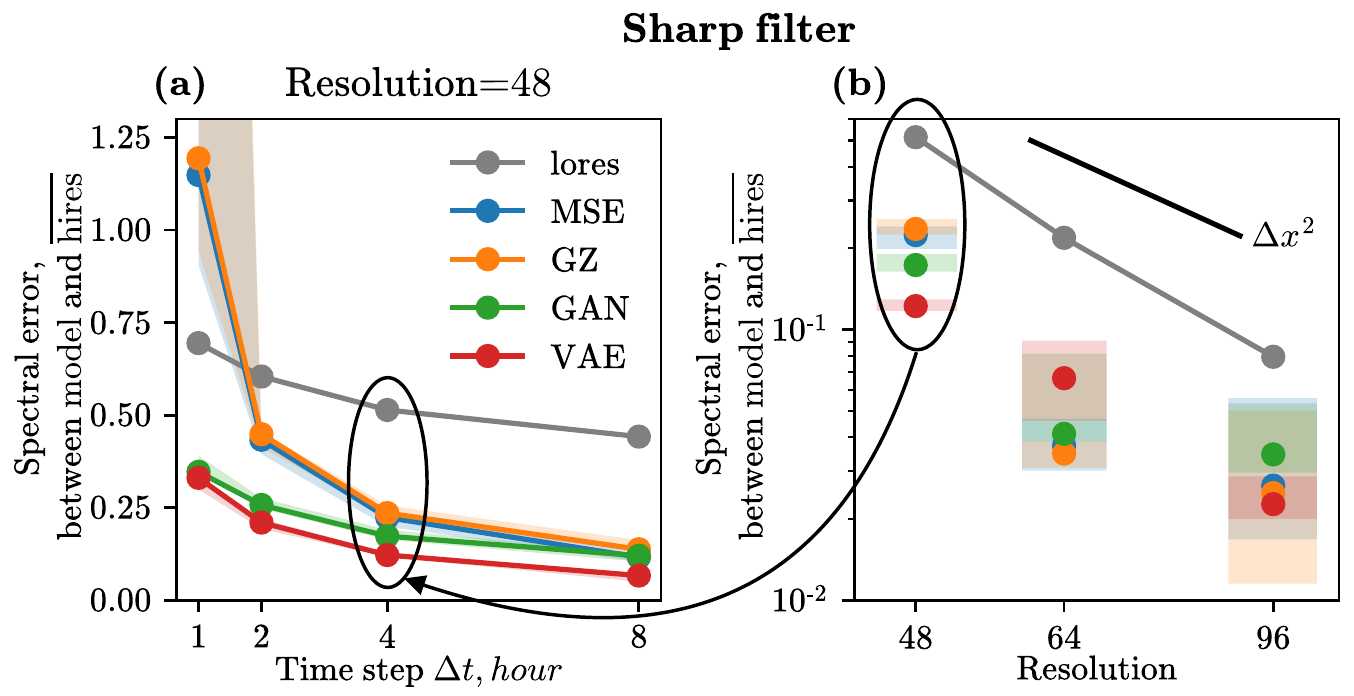}
\caption{Online metric similar to Figure \ref{fig:distribution_error}, but for spectral error. The shading area shows min-max values
among training realizations, and markers show median values.}
\label{fig:spectral_error}
\end{figure}

 We define distance between two isotropic spectra $\mathcal{E}_1, \mathcal{E}_2$ as:
\begin{equation}
    L_2(\mathcal{E}_1,\mathcal{E}_2) = \sqrt{\frac{1}{\kappa_c}\int_0^{\kappa_c} (\mathcal{E}_1(\kappa) - \mathcal{E}_2(\kappa))^2 d \kappa},
\end{equation}
where $\kappa_c$ is the truncation wavenumber of the exponential filter (Eq. \eqref{eq:model_filter}). And average normalized distance between model on a coarse grid ("model"{}) and filtered and coarsegrained high resolution simulation ("$\overline{\mathrm{hires}}$")
\begin{equation}
    \frac{L_2(\mathcal{E}_{\mathrm{model}}, \mathcal{E}_{\overline{\mathrm{hires}}})}{L_2(0, \mathcal{E}_{\overline{\mathrm{hires}}})}
\end{equation}
over "energy transfer"{}, "energy source"{} and kinetic energy spectrum in upper and lower fluid layers. The described spectral error is shown in Figure \ref{fig:spectral_error}.

\begin{figure} [h!]
\centering
\includegraphics[width=1.0\textwidth]{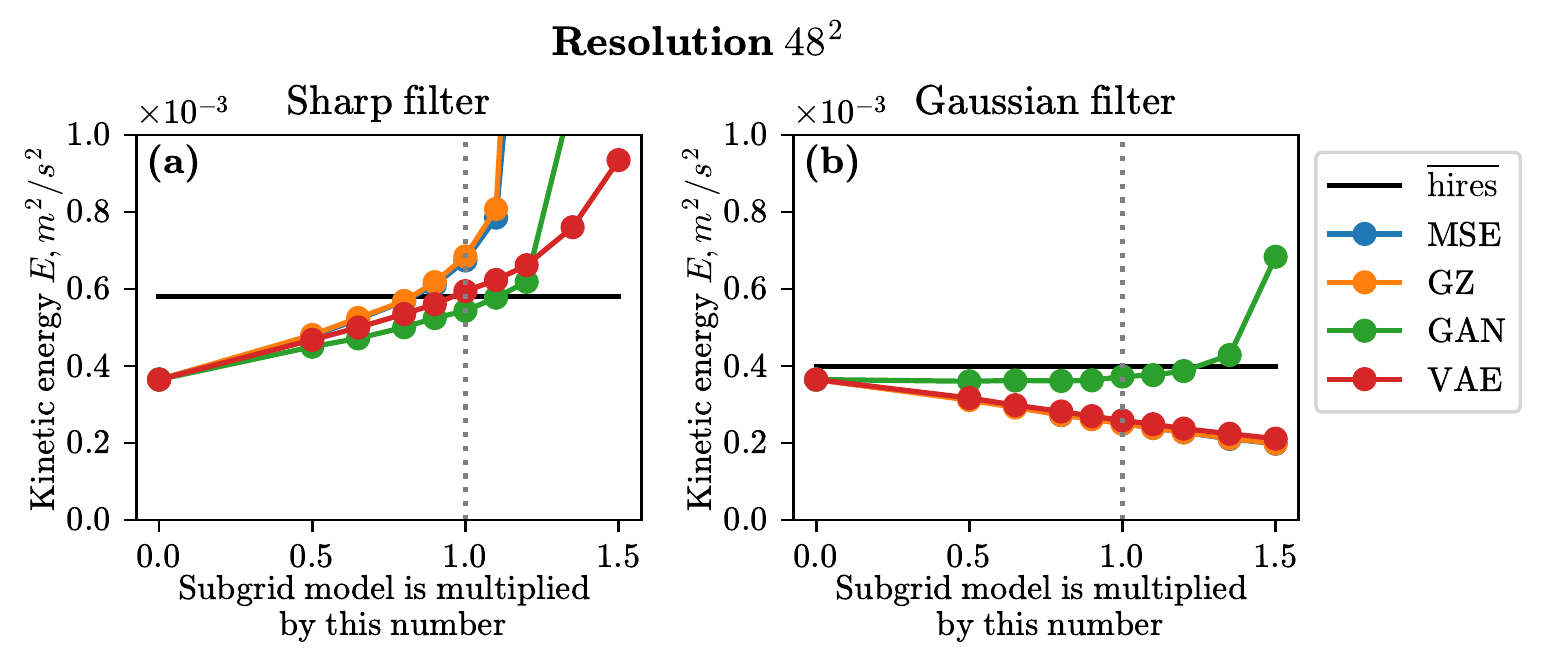}
\caption{We multiply the subgrid model by a parameter $\alpha \in [0,1.5]$ as  $\widetilde{S}\to \alpha \widetilde{S}$ and show the kinetic energy after spin-up. Subgrid models which efficiently simulate backscatter are able to energize the flow when the amplitude is increased, see supplemental Figure S9 in \citeA{zanna2020data}. All models trained for the Sharp filter efficiently energize the flow, but for the Gaussian filter they mostly do not energize the flow. Time step $\Delta t$ is 4 hours.}
\label{fig:amp_sensitivity}
\end{figure}

\begin{figure} [h!]
\centering
\includegraphics[width=1\textwidth]{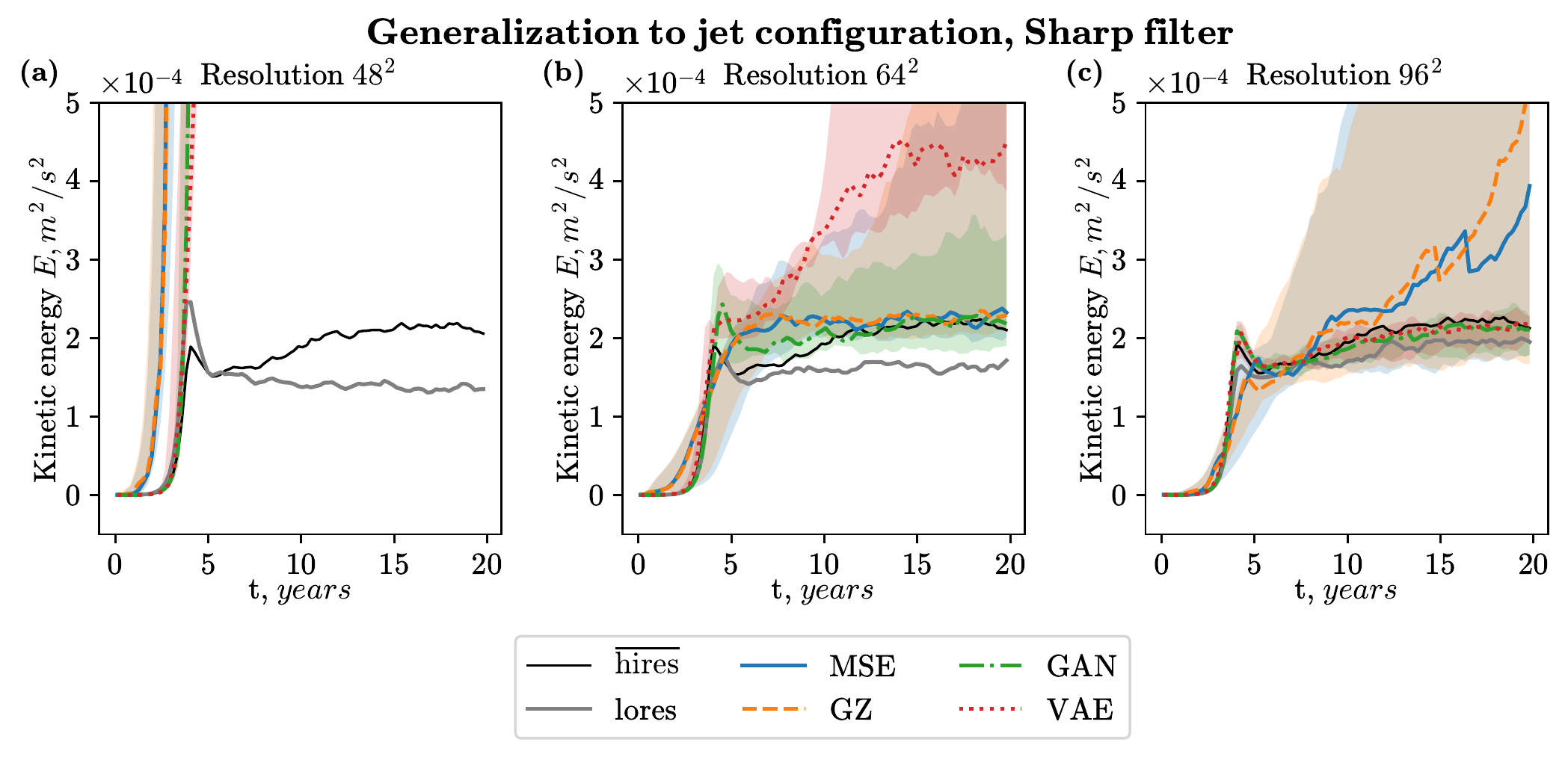}
\caption{Online generalization to turbulence configuration with jets \cite{ross2022benchmarking}. Generative models (GAN and VAE) clearly better reproduce transitional flow ($t < 2$ years), however many of the presented models have problems with numerical stability at a larger time. Improving the generalization capabilities of presented models requires further research. The shading area shows min-max values among training realizations, lines show median value. The time step is 2 hours.}
\label{fig:generalization_jet_online}
\end{figure}

Figure \ref{fig:amp_sensitivity} shows the sensitivity to the amplitude of the parameterization and 
Figure \ref{fig:generalization_jet_online} shows the online generalization to the jet dataset. Table \ref{tab:runtime} shows runtime of the parameterized models.

\begin{table}[]
\begin{tabular}{l|c|c|ccccc}
$\Delta t$      & 1 hour       & 2 hour      & \multicolumn{5}{c}{4 hour}                                   \\
\hline
$n\times n$     & $256 \times 256$          & $96 \times 96$          & \multicolumn{5}{c}{$48 \times 48$}                                       \\
\hline
Model   & --            & --           & --      & MSE     & GZ     & GAN          & VAE          \\
Runtime, sec & 1300 & 130 & 25.4 & 756 & 1480 & 784 & 782
\end{tabular}
\caption{Runtime on one CPU core for unparameterized model ("--"{}) and ML-based parameterizations to integrate QG model in time for 20 years. Theoretically, we expect that the runtime for MSE, GAN, and VAE models should be the same, and for GZ is twice as large. Runtime for the GZ model can be reduced if aggregate mean and variance channels into one CNN network, as it is done in \citeA{guillaumin2021stochastic}.}
\label{tab:runtime}
\end{table}

\clearpage

\section*{Data Availability Statement}
Python software used for training and evaluation of the subgrid models is available via \url{https://github.com/m2lines/pyqg_generative} (see the archived version on Zenodo, \citeA{pavel_perezhogin_2023_7641961}). We provide training and simulation data on Zenodo \cite{pavel_perezhogin_2023_7622683}.

\acknowledgments

This research is supported by the generosity of Eric and Wendy Schmidt by recommendation of Schmidt Futures, as part of its Virtual Earth System Research Institute (VESRI).
C.F.G. was partially supported by NSF DMS Grant 2009752. This research was also supported in part through the NYU IT High Performance Computing resources, services, and staff expertise and by the National Science Foundation under Grant No. NSF PHY-1748958.
The authors would like to thank the members of M$^2$LInES for their helpful comments and discussions.

% This section is optional. Include any Acknowledgments here.
% The acknowledgments should list:\\
% All funding sources related to this work from all authors\\
% Any real or perceived financial conflicts of interests for any author\\
% Other affiliations for any author that may be perceived as having a conflict of interest with respect to the results of this paper.\\
% It is also the appropriate place to thank colleagues and other contributors. AGU does not normally allow dedications.

%% ------------------------------------------------------------------------ %%
%% References and Citations

%%%%%%%%%%%%%%%%%%%%%%%%%%%%%%%%%%%%%%%%%%%%%%%
%
\bibliography{agusample}
%
% don't specify bibliographystyle

% In the References section, cite the data/software described in the Availability Statement (this includes primary and processed data used for your research). For details on data/software citation as well as examples, see the Data & Software Citation section of the Data & Software for Authors guidance
% https://www.agu.org/Publish-with-AGU/Publish/Author-Resources/Data-and-Software-for-Authors#citation

%%%%%%%%%%%%%%%%%%%%%%%%%%%%%%%%%%%%%%%%%%%%%%%

%\bibliography{enter your bibtex bibliography filename here}

%Reference citation instructions and examples:
%
% Please use ONLY \cite and \citeA for reference citations.
% \cite for parenthetical references
% ...as shown in recent studies (Simpson et al., 2019)
% \citeA for in-text citations
% ...Simpson et al. (2019) have shown...
%
%
%...as shown by \citeA{jskilby}.
%...as shown by \citeA{lewin76}, \citeA{carson86}, \citeA{bartoldy02}, and \citeA{rinaldi03}.
%...has been shown \cite{jskilbye}.
%...has been shown \cite{lewin76,carson86,bartoldy02,rinaldi03}.
%... \cite <i.e.>[]{lewin76,carson86,bartoldy02,rinaldi03}.
%...has been shown by \cite <e.g.,>[and others]{lewin76}.
%
% apacite uses < > for prenotes and [ ] for postnotes
% DO NOT use other cite commands (e.g., \citet, \citep, \citeyear, \citealp, etc.).
% \nocite is okay to use to add references from your Supporting Information
%

\end{document}